\documentstyle[jkas35]{article}
\runningauthor{LEE}
\runningtitle{Elliptical Galaxy formation}
\beginpage{1}
\endpage{99}

\begin{document}

\title{On The Formation of Giant Elliptical Galaxies and Globular Clusters}

\author{Myung Gyoon Lee}
%\footnote{Visiting Investigator, Department of Terrestrial Magnetism, Carnegie Institution of Washington}}

\address{Astronomy Program, SEES,
Seoul National University, Seoul 151-742, Korea \\
{\it E-mail: mglee@astrog.snu.ac.kr}}

%\address{\normalsize{\it (Received mmm, dd, 2003; Accepted ???. ??,2003)}}
\address{\normalsize{\it to appear in the Journal of Korean Astronomical Society, 2003}}

\abstract{
I review the current status of understanding when, how long, and how giant elliptical galaxies formed,
focusing on the globular clusters.
Several observational evidences show that massive elliptical galaxies formed at $z>2$ ($>10$ Gyr
ago). Giant elliptical galaxies show mostly a bimodal color distribution of globular clusters,
indicating a factor of $\approx 20$ metallicity difference between the two peaks.
The red globular clusters (RGCs) are closely related with the stellar halo in color and spatial
distribution, while the blue globular clusters (BGCs) are not.
The ratio of the number of the RGCs and that of the BGCs varies depending on galaxies.
It is concluded that the BGCs might have formed 12--13 Gyr ago,
while the RGCs and giant elliptical galaxies might have formed similarly 10-11 Gyr ago.
It remains now to explain the existence of a gap between the RGC formation epoch and the BGC
formation epoch,
and the rapid metallicity increase during the gap ($\Delta t \approx 2$ Gyr).
If hierarchical merging can form a significant number of giant elliptical galaxies
$>10$ Gyr ago, several observational constraints from stars and globular clusters in
elliptical galaxies can be explained.
}

\keywords{galaxies:general --- galaxies:formation --- galaxies: globular clusters ---
galaxies: elliptical}
\maketitle

\section{INTRODUCTION:UNSOLVED MYSTERY}

\begin{figure*}
   \begin{minipage}{\textwidth}
    \begin{center}
    \leavevmode
    \epsfxsize = 16.0cm
    \epsfysize = 16.0cm
    \epsffile{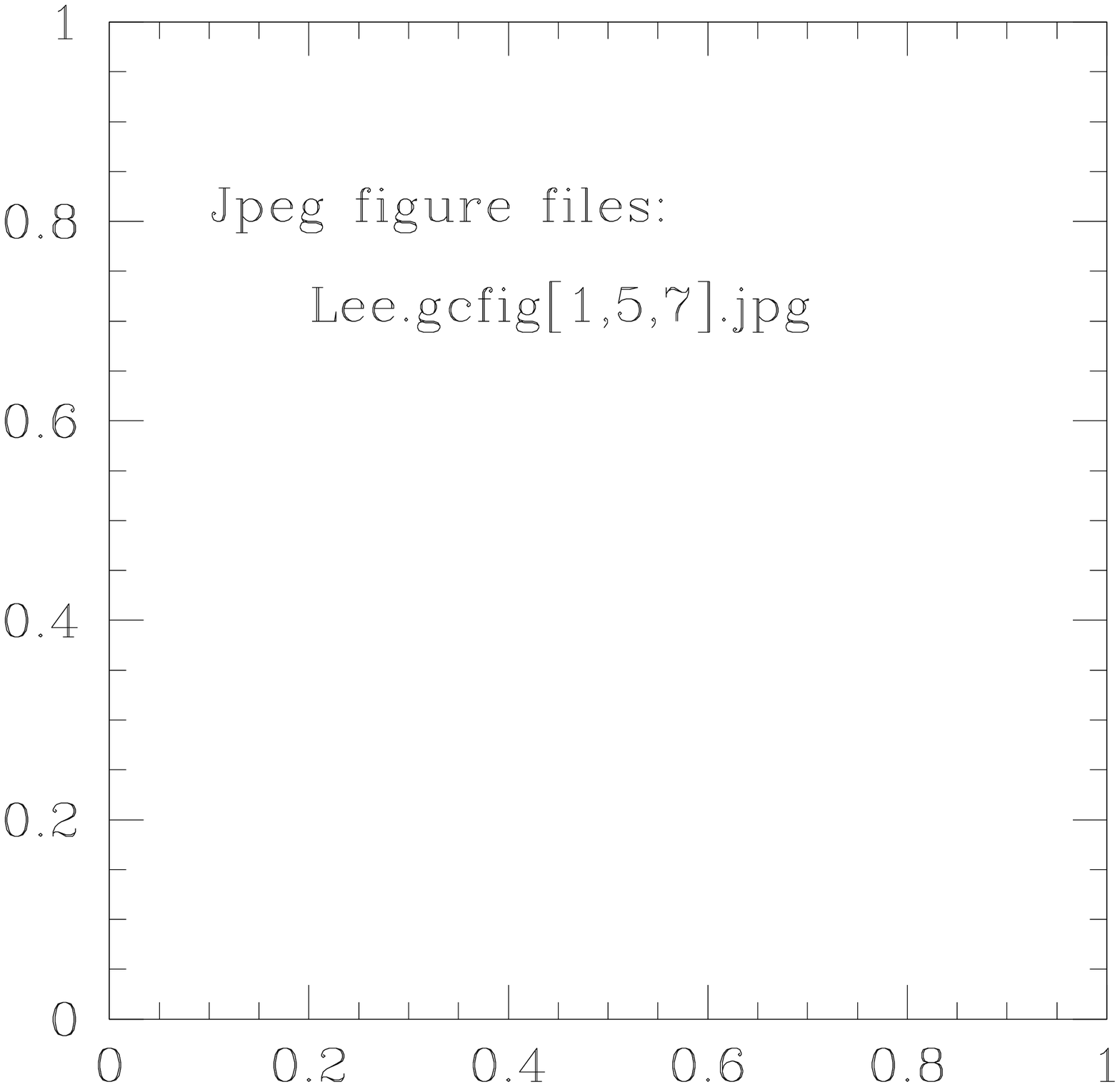} %Lee.gcfig1.ps} %m49galaxy.ps}
  \end{center}
\caption{A color image of M49, the brightest galaxy (gE) in the Virgo cluster, 
created
by combining $C$ and $T_1$ images taken at the KPNO 4m.
The field of view is $16'\times 16'$. A fraction of stellar halo light of M49 
was subtracted
from the original image to show better the globular clusters.
Most of the faint point sources are globular clusters. Several dwarf elliptical
galaxies are also seen around M49.
One blue feature which looks like an anchor in the south-east of M49 is UGC 7636, 
a dwarf irregular galaxy interacting with M49.}
\label{fig1}
\end{minipage}
\end{figure*}

Galaxies are a gateway to understanding the formation and evolution of large-scale structures
in the universe.
Morphological types of galaxies are diverse from dwarf spheroidal galaxies to spiral galaxies with
beautiful arms,
and masses of galaxies span a very large range from $10^7 M_\odot$ of dwarf galaxies
to $10^{12}$ $M_\odot$ of giant elliptical galaxies.
Here I will concentrate only on giant elliptical galaxies ($M_V<-20$ mag)
which are often found
in the centers of rich clusters.
% (of course, other types of galaxies are important as well).

Globular clusters (with masses of $10^3$ to $10^6$ $M_\odot$) are found  in many of these galaxies
(Ashman \& Zepf 1998).
The brightest old globular clusters ($M_V\approx -10$ mag) are brighter than the lowest-mass dwarf galaxies,
but the globular clusters (with half mass radii of order 2--10 pc) are incomparably smaller
than dwarf galaxies  which are larger than a few 100 pc.
Dwarf galaxies host globular clusters, but globular clusters cannot host dwarf galaxies.

Among the diverse galaxies,
elliptical galaxies (and the bulges of spiral galaxies as well) share some common properties
with globular clusters.
At first look both look monotonously simple (elliptical or globular shape in uniform color in the images). This leads
us to consider that they must be composed of one single old stellar population,
and to conclude that they may be the first stellar systems which formed in the universe.
That is an old picture of elliptical galaxies and globular clusters we had in the past before
modern data came out.

The history changed in the 1990's with the advent of the Hubble Space Telescope, large-format
CCD cameras and large ground-based telescopes. It turned out that the elliptical galaxies and globular
clusters are not that simple.
Two of the most noteworthy findings related to globular clusters in elliptical galaxies
are 1) that the color (and metallicity) distribution of globular clusters in elliptical galaxies
is often bimodal (e.g., M87: Lee \& Geisler 1993, Whitmore  et al. 1995, Elson \& Santiago 1996;
M49: Geisler, Lee, \& Kim 1996),
and 2) that there exist many blue bright clusters
in the interacting/merging systems of galaxies that are much brighter
than the Galactic halo globular clusters. These can be progenitors of globular clusters
(e.g., Whitmore \& Schweizer 1995, Whitmore 2000, Schweizer 2002a,b).
These findings showed that these simple-looking stellar systems are complex in reality,
and that globular clusters are forming in the process of galaxy merging,
providing critical constraints to models of galaxy formation.

When, how long, and how these galaxies and globular clusters formed
(formation epoch, formation duration, and formation process) is a key question in modern
cosmology. It is a long-standing question.
Models of formation of these objects came out as early as in 1960's and 1970's
(Eggen, Lynden-Bell, \& Sandage 1962, Partridge \& Peebles 1967, Peebles \& Dicke 1968,
Tinsley 1972, Larsen 1974, Toomre 1977), and
models based on numerical simulations with high resolution are coming out today
(e.g., Steinmetz \& Navarro 2002, Bekki et al. 2002, Meza et al. 2003, Beasley et al. 2003, Kravtsov \& Gnedin 2003).
In spite of many efforts for this problem over many years,
this problem is still an unsolved mystery.

Most previous work on the formation of elliptical galaxies
is based on integrated properties of unresolved stars in the galaxies.
However, globular clusters play a significant
role in our understanding how elliptical galaxies formed.
There was only limited observational information of globular clusters in  elliptical galaxies
in the past, but new observational information about them is flooding these days,
making modelers busy.
It is necessary to use both stars and globular clusters
to solve the unknown mystery of the formation of elliptical galaxies.

Figure 1 shows an example of a giant elliptical galaxy, M49, with its globular clusters.
%Figure 1 displays a $16'\times 16'$ image of
M49 (NGC 4472) %which
is the brightest galaxy (E2/S0) in the Virgo cluster which
is at the distance of about 15 Mpc. The image was created by combining $C$ and $T_1$ images
obtained at the KPNO 4m. A significant fraction of stellar halo light of M49 was subtracted
from the original images to show better the point sources in the image.
Note the numerous point sources concentrated around the center of M49,
most of which are genuine globular clusters.
This figure shows that those globular clusters can be a critical tool
to investigate the nature of elliptical galaxies.
In Figure 1 one blue feature that looks like an anchor in the south-east of M49 is UGC 7636,
a dwarf irregular galaxy interacting with M49.
Several young clusters are found in this region and the color of this galaxy is blue,
showing that clusters form during the interaction of galaxies (Lee, Kim, \& Geisler 1997).

There are numerous studies and reviews on the formation of elliptical galaxies and globular
clusters (e.g., Burkert 1994, Renzini 1999, Whitmore 2000,
Silk \& Bouwens 2001, Ellis 2001, Harris 2001,2003, Gnedin, Lahav, \& Rees 2001, Gnedin 2002,
Peebles 2002, Matteucci 2002, Steinmetz 2002, Conselice 2002, C\^ote 2002, Burkert \& Naab 2003).
A recent conference proceeding related to this topic is in Geisler et al. (2002).
%and more coming out soon. %Piotto et al. (2002), and Kissler-Patig et al. (2002).
Even reviewing of reviews is not easy.
Here I review the current status of understanding the formation of elliptical galaxies
(mostly giant elliptical galaxies with $M>10^{11}~M_\odot$), focusing on globular clusters.

This paper is composed as follows.
Section II points out key elements for solving the key question of galaxy formation,
and Section III describes briefly two proto-models of formation of elliptical galaxies.
Observational constraints from stars in elliptical galaxies are listed in Section IV, and
observational constraints from globular clusters in elliptical galaxies are given in Section V.
Section VI describes briefly models of formation of globular clusters.
The final section concludes with the present status,
and presents a sketch for the formation of giant
elliptical galaxies.

\section{KEY ELEMENTS FOR UNSOLVED MYSTERY}

Primary goals of studying the formation of structures (dust, planets, solar systems,
stars, star clusters, galaxies, galaxy clusters, etc) are to find when, how long, and how they
formed. These goals are not easy to achieve even in the case of stars (see Elmergreen 2002).
In general it is often easy(?) to figure out possible ways to form structures,
and `when' (formation epoch) and `how long' (formation duration) are needed to constrain
the possible `how's (formation processes).

However, `when' and `how long' are often difficult to know accurately and precisely,
especially when the targets are old and far away, like elliptical galaxies.
With the exception of NGC 5128, %The nearest giant elliptical galaxy, NGC 5128, is
at the distance of about 4 Mpc,
all of the giant elliptical galaxies are located beyond 10 Mpc.
Most of the globular clusters in the universe may belong to these giant elliptical galaxies,
and they must contain critical information on the formation of elliptical galaxies
(although they occupy only a tiny fraction of mass in galaxies).

\begin{figure*}
  \begin{minipage}{\textwidth}
  \begin{center}
    \leavevmode
   \epsfxsize = 16.0cm
   \epsfysize = 16.0cm
    \epsffile{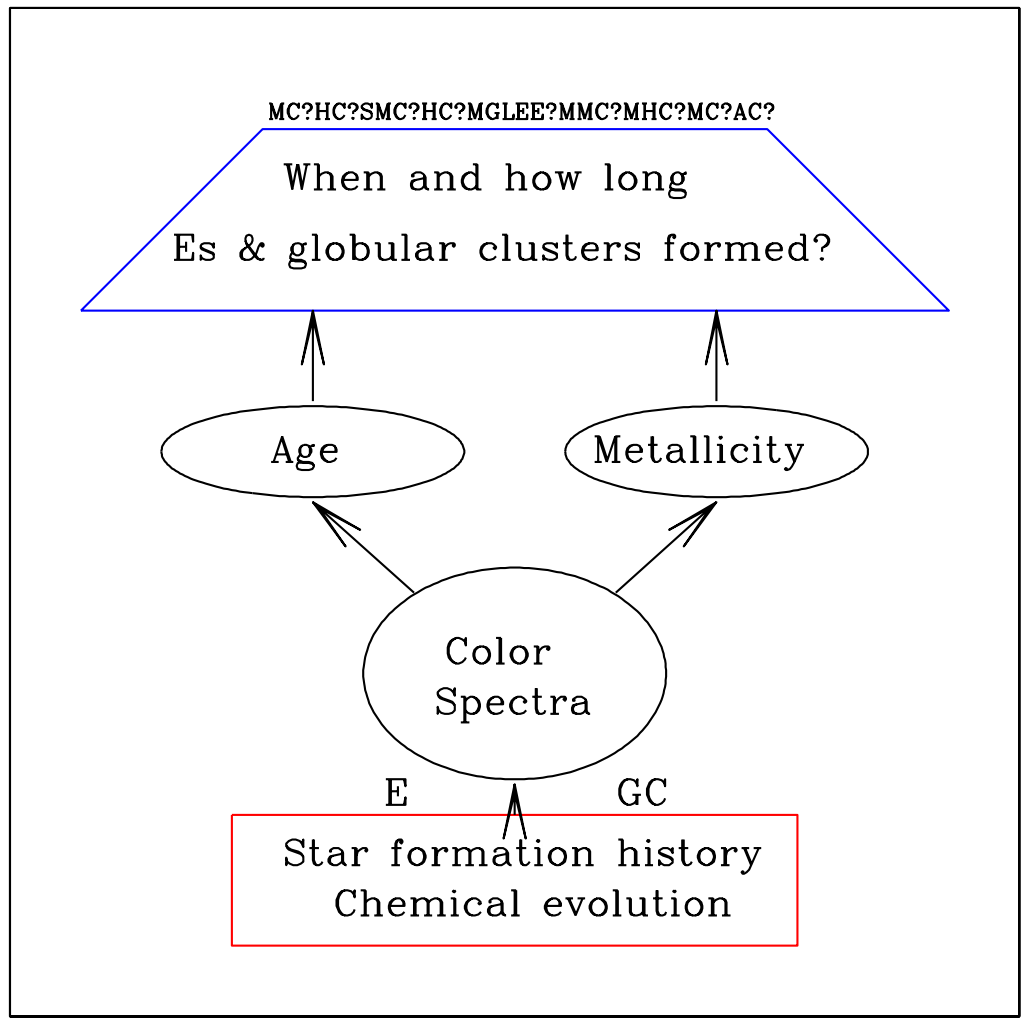} %Eform1.ps}
  \end{center}
  \vskip -1.8cm
\caption{A schematic diagram showing the key elements in the study
of formation of elliptical galaxies and globular clusters.
Photometric colors and spectral features in spectra are observational tools
to figure out star formation history and chemical evolution in  elliptical galaxies and globular clusters.
Integrated colors and spectra are obtained from the mixed populations in these
stellar systems,
and are dependent on  both age and metallicity.
A core of the problem is in breaking the age-metallicity degeneracy in colors
and spectra to understand when and how long these objects formed.}
\label{fig2}
\end{minipage}
\end{figure*}

Figure 2 displays a schematic diagram showing the key elements in the study of formation of
elliptical galaxies and globular clusters.
Galaxies form and evolve chemically (photometrically and dynamically as well),
and we need to know the star formation history and chemical enrichment history of these galaxies.

Two major observational tools obtained for distant galaxies and globular clusters are
photometric color and/or spectral features.  These two pieces of information are based on
the integrated light from the various kinds of stellar populations.
Spectra contain more information than color, but colors are easier to get than spectra.
So the information we have now is color for most objects and spectra for a small number.

A well-known problem in using integrated color is
that it is difficult to distinguish between the metal-poor old systems and
the metal-rich intermediate-age systems, because age and metallicity have similar effects
to color (and age and metallicity are often correlated).
Increasing age and/or metallicity makes the color redder.
Modern history in the studies of elliptical galaxies and globular clusters is full of
efforts to break this age-metallicity degeneracy.
Spectral information is very important in breaking this age-metallicity degeneracy,
but not yet enough to do it.

%Once we measure the age of stellar systems and its range, then we can tell
%when and how long elliptical galaxies and globular cluster formed.
So the problem of knowing `when' and `how long' is observationally equivalent
to the problem of breaking the age-metallicity degeneracy.
It is easy in principle, but difficult in reality.
However, we are getting closer to the goal.

One of the nagging factors in these studies is large uncertainties of the parameters used
for conversion between the lookback time and redshift. Astronomers working with age-dating of
stellar systems almost always used time,
while astronomers working with distant objects always used redshift $z$.
So it is not easy to compare directly results given in terms of time and $z$.

The lookback time is given as a function of redshift,$z$, as follows (Hogg 1999):

\begin{displaymath}
t_{L}(z) = {9.78 \over {h}} \times
~~~~~~~~~~~~~~~~~~~~~~~~~~~~~~~~~~~~~~~~~~~~~~~~\\
\end{displaymath}
\begin{displaymath}
~~~~~~~~~ \int^z_0  {dz' \over
{ (1+z') \sqrt{ \Omega_M (1+z')^3+\Omega_k (1+z')^2+\Omega_{\Lambda} } } }
\end{displaymath}

 \noindent where $h$ is a normalized Hubble constant, $H_0/100$ [km~ s$^{-1}$~ Mpc$^{-1}$], and
 $\Omega_M$, $\Omega_{\Lambda}$, and $\Omega_k$ are, respectively,
 density parameters for matter(including dark matter and baryonic matter), dark energy, and curvature.

Now several independent sources provide converging values for these parameters so the direct comparison
between the time and $z$ is possible.
Spergel et al. (2003) provide precise estimates of cosmological parameters
combining WMAP results on microwave cosmic background radiation, Supernovae Ia, and large-scale
galaxy redshifts: $h=0.71^{+0.04}_{-0.03}$, $\Omega_b h^2 = 0.0224\pm0.0009$,
$\Omega_M h^2 = 0.135^{+0.008}_{-0.009}$, $\Omega_{total}=1.02\pm0.02$,
the equation of state of the dark energy $w<-0.78$, and
the age of the universe, $13.7\pm0.2$ Gyr.
Tonry et al. (2003) presented their estimates of some of these parameters based on
high-$z$ Supernovae Ia results, which are consistent with Spergel et al. (2003)'s.
Spergel et al.'s age estimate
$13.7\pm0.2$ Gyr is only slightly larger than the mean age of the 17 metal-poor globular clusters
in the halo of our Galaxy, 12.6 Gyr with a 95\% confidence level lower limit of 11.2 Gyr
(Krauss \& Chaboyer 2003).
According to the WMAP results, the reionization epoch may be $z_r=17\pm3$, or
$11<z_r<30$ (corresponding to times $100<t_r <400$ Myr after the Big Bang) (Kogut et al. b2003).
We adopt
a current standard model of cosmology, concordance model  to calculate the lookback time:
$\Omega_M=0.3$, $\Omega_{\Lambda}=0.7$, $\Omega_{k}=0.0$, and $h=0.71$.

\section{MODELS OF FORMATION OF ELLIPTICAL GALAXIES}

The formation of elliptical galaxies may be closely related with the formation of
bulges and halos in spiral galaxies. So the models for the elliptical galaxies are also
related with the models for the bulges and halos in spiral galaxies. One motivates the other.
Historically the models of elliptical galaxies followed the models of the halo of  our Galaxy,
a spiral galaxy.
There are two main competing models to explain the formation of  elliptical galaxies:
monolithic collapse model and hierarchical merging model.

\subsection{Monolithic Collapse Models}

The concept of monolithic collapse model (MCM) was introduced first
by Eggen, Lynden-Bell, \& Sandage
(1962) who tried to explain the radial orbits of the stars in the halo of our Galaxy.
This model was expanded to the case of elliptical galaxies later (Partridge \& Peebles 1967,
Tinsley 1972, Larson 1974, Chiosi \& Carraro 2002, Matteucci 2002, Kawata \& Gibson 2003).
In this model,  elliptical galaxies (and the spheroidal components of disk galaxies as well)
formed as a result of large protogalactic clouds collapsing and forming stars
in a very short time of about $10^7 \sim 10^8$ yr very early (at $z>3$).
After the violent burst of star formation leading to high metallicity very early,
these galaxies evolved quiescently until today (passive evolution).

\subsection{Hierarchical Merging Models}

Toomre (1977) proposed an alternative model where elliptical galaxies formed when two spiral
galaxies mer\-ged, and Searle \& Zinn (1978) suggested that the halo of our Galaxy might have formed
via accreting transient proto-Galactic fragments with dwarf galaxy-mass for an extended period of time.
These ideas were developed into hierarchical merging (or clustering) models (HMM)
in the context of cold dark matter (CDM) model of structure formation
where small objects are progressively
incorporated into larger structures, and stars form at the baryonic cores
embedded in the dark matter halo
(Kauffmann, White, \& Guiderdoni 1993, Benson, Ellis, \& Menanteau 2002, Benson \& Madau 2003,
Steinmetz \& Navarro 2003, Helly et al. 2003, Khochfar \& Burkert 2003, Meza et al. 2003).
In this model, massive elliptical galaxies formed much later and much longer
than those monolithic collapse models predict.
Merging process can  be roughly divided in two: major merger (merging of  large galaxies)
and minor merger (accretion/tidal stripping of low-mass systems like dwarf galaxies, globular clusters,
and dwarf protogalactic clouds).
These models predict that about half of the massive elliptical galaxies formed at $z<1$ (Peebles 2002).

\section{OBSERVATIONAL CONSTRAINTS\\ FROM STARS}

Elliptical galaxies show two kinds of integrated properties of stars:
one that almost all elliptical galaxies
share (uniformity and family) and the other that only some elliptical galaxies show (peculiarity).
The former includes the color-magnitude relation, fundamental planes, color gradients, and surface
brightness profiles, and the latter includes morphological and kinematical peculiar features
(ripples, shells, and kinematically decoupled cores).
These are used as constraints of models of formation of elliptical galaxies
(e.g., Bernardi et al. 2003a,b).

\subsection{Structure and Rotation}

The surface brightness profiles of elliptical galaxies (and the bulges of disk galaxies)
follow generally a de Vaucouleurs $r^{1/4}$ law
(except for disky elliptical galaxies which follow exponential law).
Elliptical galaxies are slow rotators, and are pressure-supported (dynamically hot), while
spiral galaxy disks are fast rotators, and rotation-supported (dynamically cold).
Giant elliptical galaxies are flattened by an anisotropic velocity dispersion,
while faint elliptical galaxies (and bulges in spiral galaxies) are in general isotropic
and rotationally flattened (Bender, Burstein, \& Faber 1992).

%These can be explained both by MCM and HMM.

\subsection{Color-Magnitude Relation}

\begin{figure*}
\begin{minipage}{\textwidth}
  \begin{center}
    \leavevmode
    \epsfxsize = 16.0cm
    \epsfysize = 16.0cm
    \epsffile{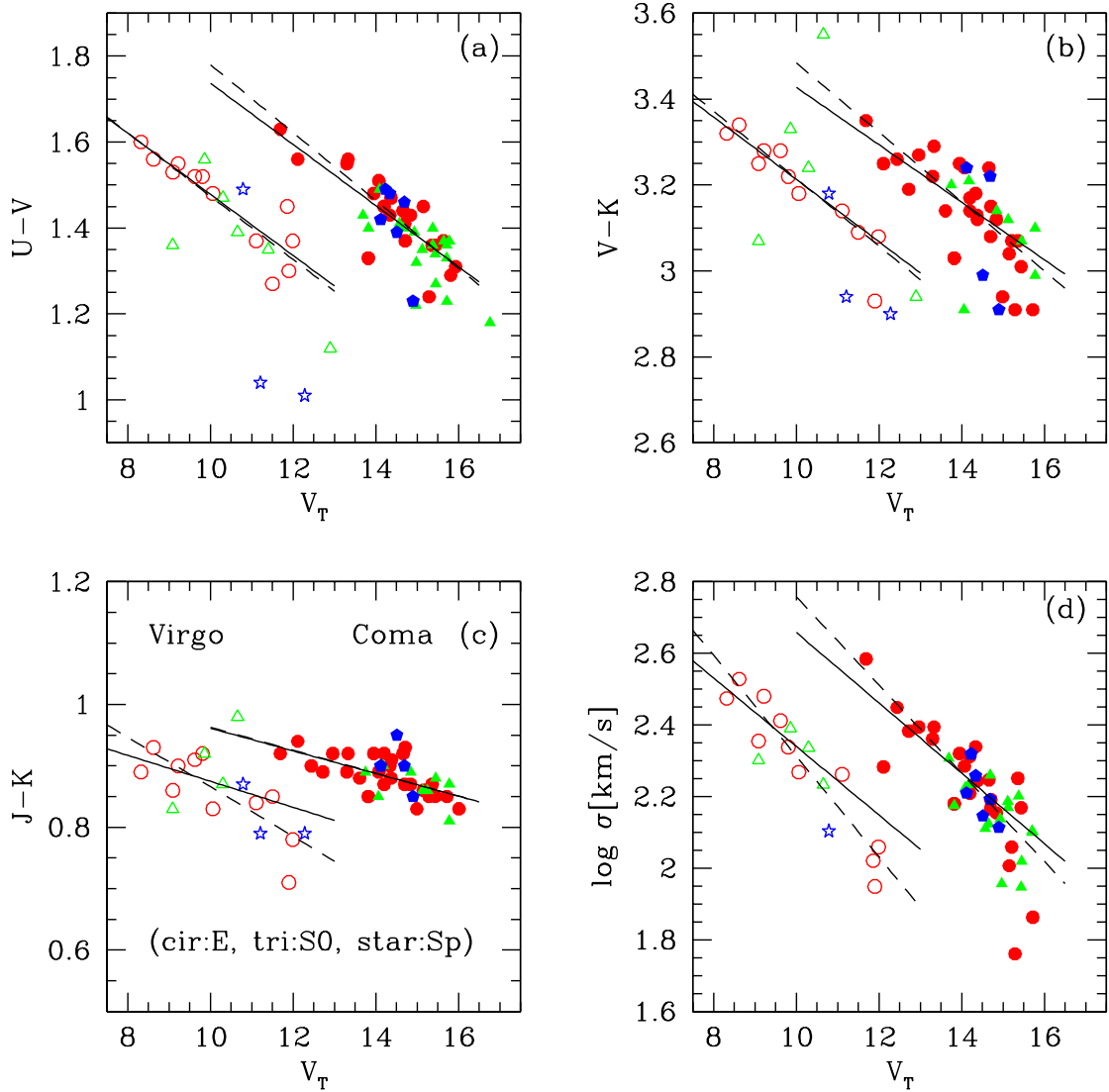} %virgocoma1.ps}
  \end{center}
\vskip -0.9cm
\caption{(a),(b), and (c) Color-magnitude relations of early type galaxies in 
Virgo (open symbols) and Coma (filled symbols)
based on the data of Bower et al. (1992). Circles, triangles and star symbols
represent, respectively, Es, S0s, and galaxies with types later than S0.
The solid lines represent the linear fits to the data of Es excluding a few 
outliers at the faint end,
and the dashed lines the fits for all galaxies given by Bower et al. (1992).
(d) Log $\sigma_v$ vs. total magnitude relation.
The scatters in color  of Es are mostly due to observational errors,
and the intrinsic dispersions of color are remarkably small ($<0.04$ mag).
(source: Bower et al. 1992)}
\label{fig3} \end{minipage}
\end{figure*}

The brighter elliptical galaxies are, the redder they are. It is called a color-magnitude relation (CMR)
of elliptical galaxies (and S0 galaxies). It has been known for long
(Baum 1959). Baum (1959) pointed out early a transition in color from Galactic globular
clusters, through dwarf elliptical, to giant elliptical galaxies (see his Figure 5).
Figure 3 shows the color-magnitude relation
of early type galaxies in Virgo and Coma given by Bower, Lucey, \& Ellis (1992).
The scatters in color of elliptical galaxies (and S0's)
in Virgo and Coma are found to be remarkably small ($\Delta(U-V)\approx 0.04$).
This scatter is related with age dispersion, $d(U-V)/dt\approx 0.02$ mag/Gyr for old age according to the
evolutionary population synthesis models. This small scatter indicates
that the formation duration of these galaxies is shorter than 2 Gyr
and the formation epoch is very early at $z>3$ (Bower, Lucey, \& Ellis 1992,
Terlevich, Caldwell, \& Bower 2001, Peebles 2002).

The color-magnitude relations of early type galaxies in more distant clusters
(up to $z<1.27$) and in the Hubble Deep Field (Williams et al. 1996)
are consistent with those for $z=0$, indicating the primary origin of the color-magnitude relation is
metallicity, not age (Kodama \& Arimoto 1997, Kodama et al. 1998,
Bower, Kodama, \& Terlevich 1998, Kodama, Bower, \& Bell 1999).
This is strong evidence supporting the MCM.
The HMM also predicts this relation, but with an assumption that elliptical galaxies merge
other elliptical galaxies of the same luminosity to keep the CMR.

\subsection{Fundamental Plane Relations}

Stellar systems can be characterized by three global physical parameters:
central velocity dispersion, effective radius, and effective surface brightness.
Fundamental planes of stellar systems are mathematical relations between these parameters.
%composed of these three parameters.
Elliptical galaxies in clusters are known to occupy very narrow regions in the fundmental planes
(Kormendy 1985, Bender, Burstein, \& Faber 1992,  Burstein et al. 1997).
The basic physics leading to the fundamental planes is believed to be
virial theorem and galaxy evolution history.

\begin{figure*}
\begin{minipage}{\textwidth}
  \begin{center}
    \leavevmode
    \epsfxsize = 16.0cm
    \epsfysize = 16.0cm
    \epsffile{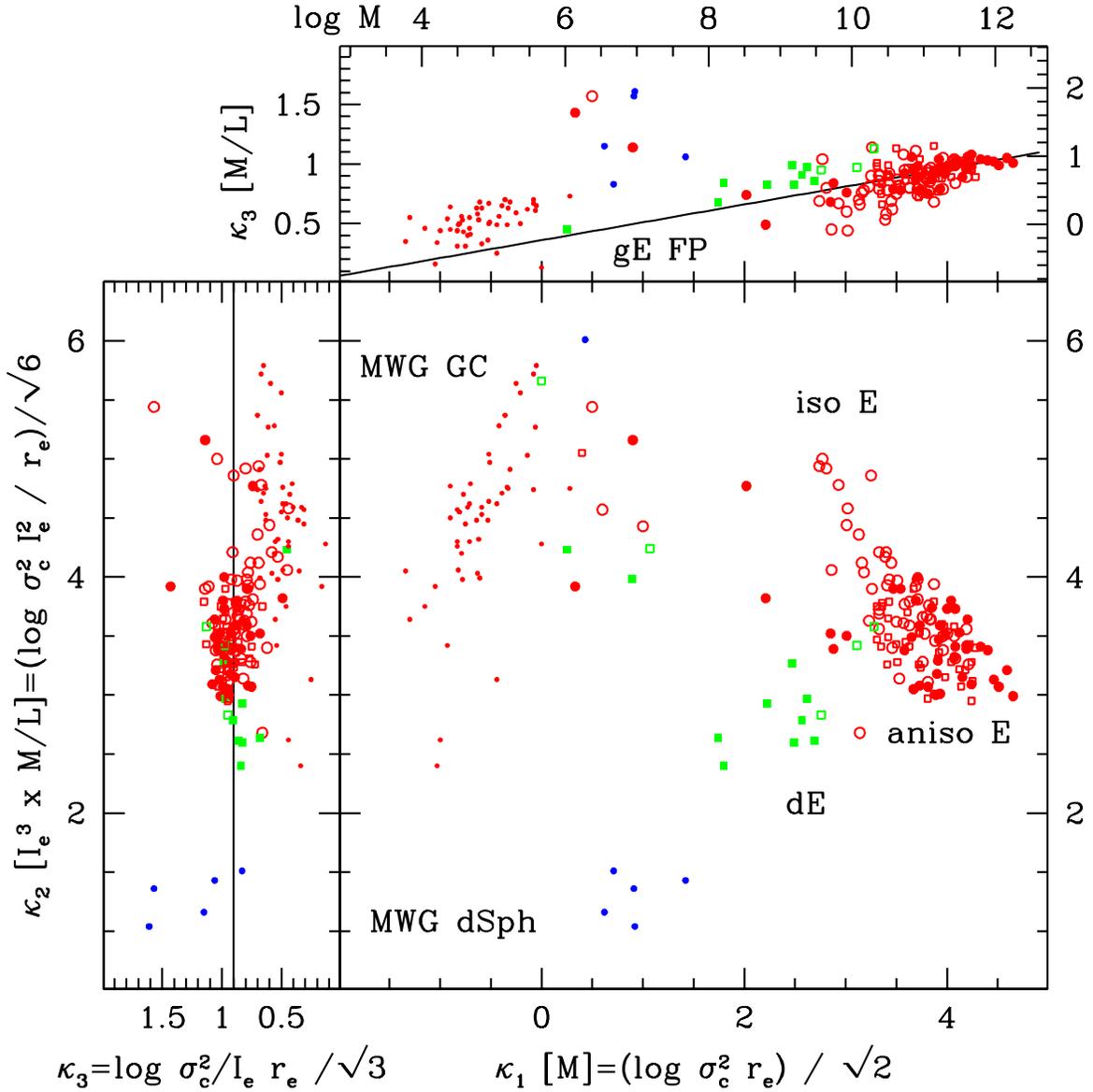} %FPEgc.ps} %burstein.fig2.ps }
  \end{center}
\vskip -0.5cm
\caption{Fundamental planes of elliptical galaxies, dwarf galaxies, and Galactic
 globular clusters %(dynamically hot galaxies)
(open circles: isotropic Es, filled circles: anisotropic Es, open squares: Es 
without isotropy information, small red dots: GCs in our Galaxy, 
small blue dots: dwarf spheroidal galaxies around our Galaxy). 
The solid line represent the fundamental planes for gEs in Virgo.
Note the tight fundamental planes of elliptical galaxies.
(source: Burstein et al. 1997)}
\label{fig4}
\end{minipage}
\end{figure*}

Figure 4 displays fundamental planes in the $\kappa$-space of elliptical galaxies,
dwarf elliptical galaxies (dE),
Local Group dwarf spheroidal galaxies (dSph), and Galactic globular clusters, using the data given
by Burstein et al. (1997).

The $\kappa$-space parameters are defined as follows:
$\kappa_1 = (\log \sigma_c^2 + \log r_e ) / \sqrt{2}$,
$\kappa_2 = (\log \sigma_c^2 + 2 \log I_e - \log r_e ) / \sqrt{6}$, and
$\kappa_3 = (\log \sigma_c^2 - \log I_e - \log r_e ) / \sqrt{3}$ (Burstein et al. 1997).

$\sigma_c^2$ is the central velocity dispersion in km s$^{-1}$, $r_e$ is effective radius in kpc,
and $I_e$ is the $B$-band surface brightness at $r_e$ in solar luminosity pc$^{-2}$
($=10^{-0.4(SB_e -27.0)}$), where $SB_e$ is the mean $B$-band surface brightness
in $B$ mag arcsec$^{-2}$ within $r_e$.
$\kappa_1$, $\kappa_2$, and  $\kappa_3$ are logarithmically related with galaxy mass (M),
surface brightness times mass-to-luminosity ratio, and mass-to-luminosity ratio (M/L).

In Figure 4, the fundamental plane in $\kappa_1 - \kappa_3$ space
defined by the gEs in Virgo and Coma is given by the solid line
($\kappa_3 = 0.15 \kappa_1 + 0.56$, $\log (M_e /L_e )= 0.184 \log M_e -1.25$).
Figure 4 shows that elliptical galaxies show very tight fundamental planes in the edge-on views,
indicating these elliptical galaxies in various environments share a common origin
(Renzini \& Ciotti 1993).
The scatters in the fundamental planes of elliptical galaxies
in clusters at $z<0.6$ lead to an estimate for formation epoch $z>2$ (Peebles 2002).
This is significant evidence supporting the MCM.
Galactic globular clusters are located in a different region in the $\kappa_1 - \kappa_2$ space,
but follow a similar trend, with offset in $\kappa_3$, with E's in the $\kappa_1 - \kappa_3$ space.

\subsection{Color Gradients}

Many bright elliptical galaxies show radial gradients of color in the sense that the color gets bluer
as the galactocentric radius increases, with
mean values of $\Delta (U-R)/\Delta log r=-0.20$ mag/dex and $\Delta (B-R)/\Delta log r=-0.09$ mag/dex
 (Peletier et al. 1990, Kim, Lee, \& Geisler 2000).
This color gradient may be due to metallicity or age variation.
Tamura et al. (2000) investigated the origin of color gradients in elliptical galaxies
by comparing the models with the sample of seven red elliptical galaxies at $z=0.1 - 1.0$ in the Hubble Deep Field.
They concluded that the color gradients in elliptical galaxies are due to metallicity effect,
not due to age effect, supporting the MCM. %monolithic collapse model of elliptical galaxies.

\subsection{High-redshift Galaxies}

Several kinds of massive objects which can be progenitors of massive elliptical galaxies
are found at high redshift $z>2$:
Lyman break galaxies, extremely red objects (ERO), submillimeter galaxies and quasars
(Giavalisco 2002, Blain et al. 2002, Cimatti et al. 2002a,b, Cimatti 2003,
Chapman et al. 2003).
This shows that massive galaxies formed at $z>2$, supporting the MCM
(see also Sarasco et al. 2002, Zirm, Dickinson, \& Dey 2003, Loeb \& Peebles 2003).
There are several pieces of evidence indicating a possibility that quasars might have formed
together with the stellar populations of early-type galaxies (Cattaneo \& Bernardi 2003). If so,
quasars might have formed after the metal-poor globular clusters.

\subsection{Morphological and Kinematical Peculiarity}

Many elliptical galaxies show peculiar and remarkable features in their morphology and kinematics:
shells, ripples, dust, jets, and kinematically decoupled cores.
These are considered as remnants of mergers (see Barnes 1998), supporting the HMM.

\subsection{Mixed Populations in the Nearest gE NGC 5128}

NGC 5128 (Cen A) is the nearest giant elliptical galaxy (at $d\approx 4$ Mpc). % and
It is an intermediate-age merger remnant (Schweizer 2002a), and
has a blue tidal stream of young stars including  a numerous young bright stars (with an age
of $\approx 350$ Myr) in the outer area (Peng et al. 2002).

In the pioneering work of resolved stars in giant elliptical
galaxies, Harris \& Harris (2002, and early references therein) presented
color-magnitude diagrams of giant stars in the halo fields of NGC 5128
(at 8 kpc, 21 kpc, and 31 kpc from the center of the galaxy center).
These data show remarkable features:
1) The metallicity (derived from $(V-I)$ colors) distribution of the stars is asymmetric and
very broad, from [Fe/H]$\approx -2.2$ dex to +0.4 dex;
2) Most of the stars are metal-rich, with a peak at [Fe/H]$\approx -0.4$ dex. The mean
metallicity of these stars is only slightly smaller than that of the metal-rich globular
clusters in the same galaxy;
3) A small fraction of stars are metal-poor.
Bekki, Harris, \& Harris (2003) and Beasley et al. (2003) used merger models
(merging of two spiral galaxies) to explain the
metallicity distribution of NGC 5128.

\subsection{Summary}

There are several constraints based on observations of stars in elliptical galaxies,
showing that massive elliptical galaxies might have formed very early at $z>2$, supporting the
MCM.
At the same time there are various observational constraints showing there were merger
events in elliptical galaxies, and the numerical simulations based on the HMM can reproduce
impressively various features of large-scale structures from dwarf galaxies to giant galaxies
and rich clusters of galaxies (see, e.g., the figures in Steinmetz 2002, Steinmetz \& Navarro 2003,
Kravtsov \& Gnedin 2003).
One critical problem in the current HMM based on the CDM models is that the HMM
cannot make many massive clusters very early
at $z>2$. Another well-known problem is that the HMM predicts too many dwarf galaxies in the
Local Group, by an order of magnitude or more than the observed (Moore et al. 1999, C\^ote, West, \& Marzke 2002).

\section{OBSERVATIONAL CONSTRAINTS\\ FROM GLOBULAR CLUSTERS}

Critical information of globular clusters in elliptical galaxies are derived from
photometry and spectroscopy.

Recent photometric studies of globular clusters in elliptical galaxies can be roughly
divided into three classes: a) $VI$ photometry of $HST$ WPFC2 for a small field often covering
the centers of the galaxies
(e.g., Lee, \& Kim 2000, Kundu \& Whitmore 2001a,b, Larsen et al. 2001),
b) wide-field multi-color photometry obtained at ground-based
telescopes (e.g., M87: Lee \& Geisler 1993, Lee et al. 2003;
M49: Geisler, Lee, \& Kim 1996, Lee, Kim, \& Geisler 1998, Rhode \& Zepf 2001;
NGC 1399: Dirsch et al. 2003),
and c) near IR photometry at the large ground-based telescopes to derive metallicity and age of
globular clusters (e.g., Kissler-Patig, Brodie, \& Minniti 2002, Puzia et al. 2002).

Recent spectroscopic studies are trying to get the metallicity, kinematics and age of  globular
clusters (e.g., Hanes et al. 2001, Cohen, Blakeslee, \& C\^ote 2003, Richtler et al. 2002,
Strader et al. 2003).
An excellent review of globular clusters in M87 and M49 in the Virgo cluster
was given by C\^ote (2002).

\subsection{Color-Magnitude Diagram}

\begin{figure*}
\begin{minipage}{\textwidth}
  \begin{center}
    \leavevmode
    \epsfxsize = 16.0cm
    \epsfysize = 16.0cm
    \epsffile{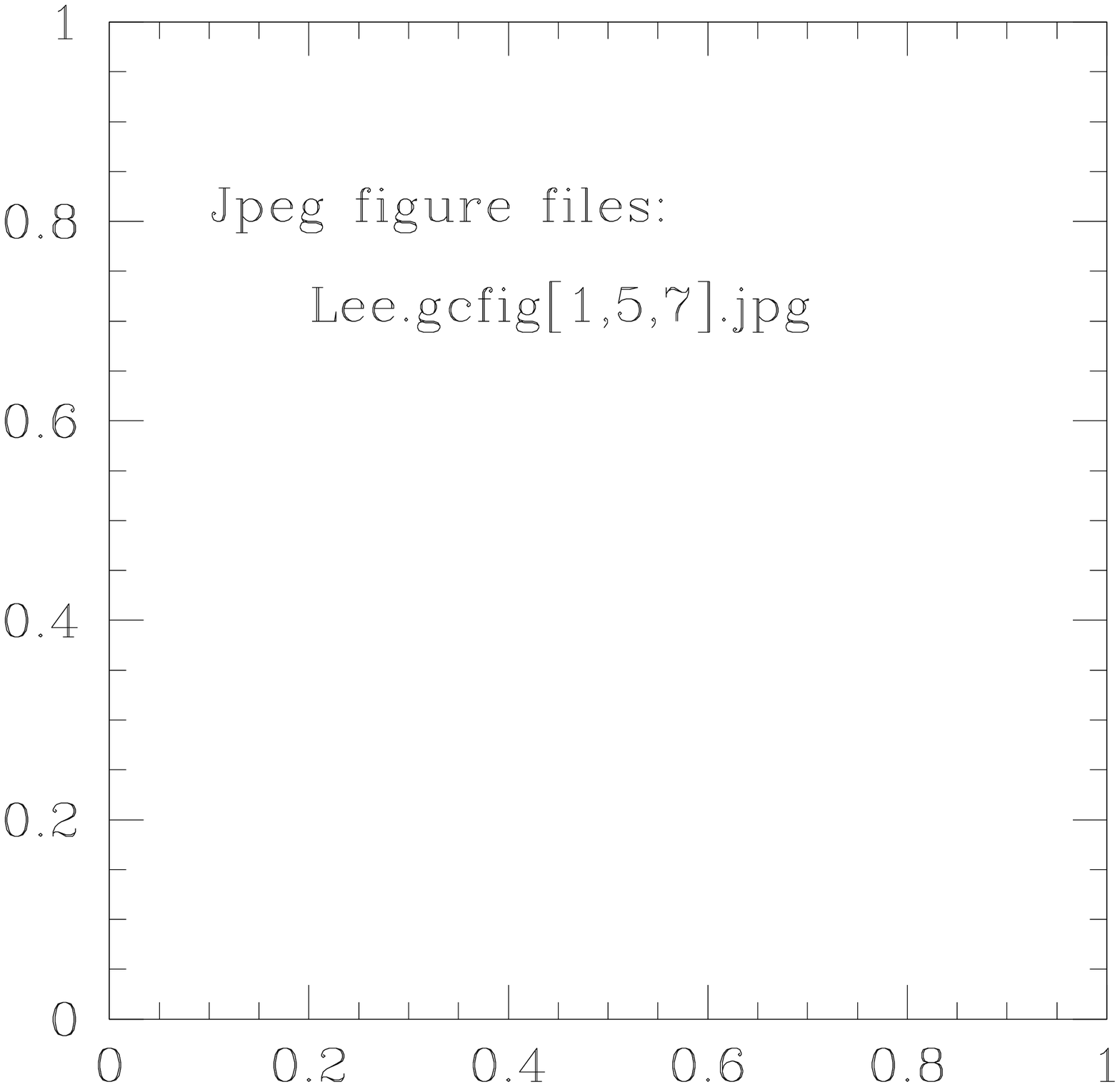} %Lee.gcfig5.ps} %virgocmd.ps} %Eform1.ps } % virgocmd.ps}
  \end{center}
\vskip -0.7cm
\caption{Color-magnitude diagrams of point sources in six giant elliptical
galaxies in Virgo.
The field of view of the images for each galaxy is $16' \times 16'$.
The order of the panels is according to the total luminosity  from the top: M49,
 M87, M60, M86, M84, and NGC 4636.
Most of the objects inside the blue boxes are probably bright globular clusters.
(source: Lee et al. 2003)}
\label{fig5}
\end{minipage}
\end{figure*}

Figure 5 displays $T_1-(C-T_1)$ color-magnitude diagrams of point sources
in six giant elliptical galaxies in the Virgo cluster (Geisler, Lee, \& Kim 1996, Lee et al. 2003).
The size of the field for each galaxy is $16' \times 16'$.
Most of the objects inside the blue boxes in Figure 5 are probably genuine globular clusters.
Most of the blue faint objects are background galaxies, and bright objects outside the
boxes are mostly foreground stars.

Figure 5 shows immediately several features of globular clusters in these galaxies:
1) there are two vertical plumes inside the boxes in most of these galaxies,
which are, respectively, blue globular clusters (BGC) and red globular clusters (RGC);
2) there is a large variation in the shape of the two plumes
 among the galaxies. A remarkable contrast is shown by M84 and NGC4636; and
3) there is a large variation in the number of globular clusters among the galaxies.

\subsection{Color Distribution}

\begin{figure*}
\begin{minipage}{\textwidth}  \begin{center}
    \leavevmode
    \epsfxsize = 16.0cm
    \epsfysize = 16.0cm
    \epsffile{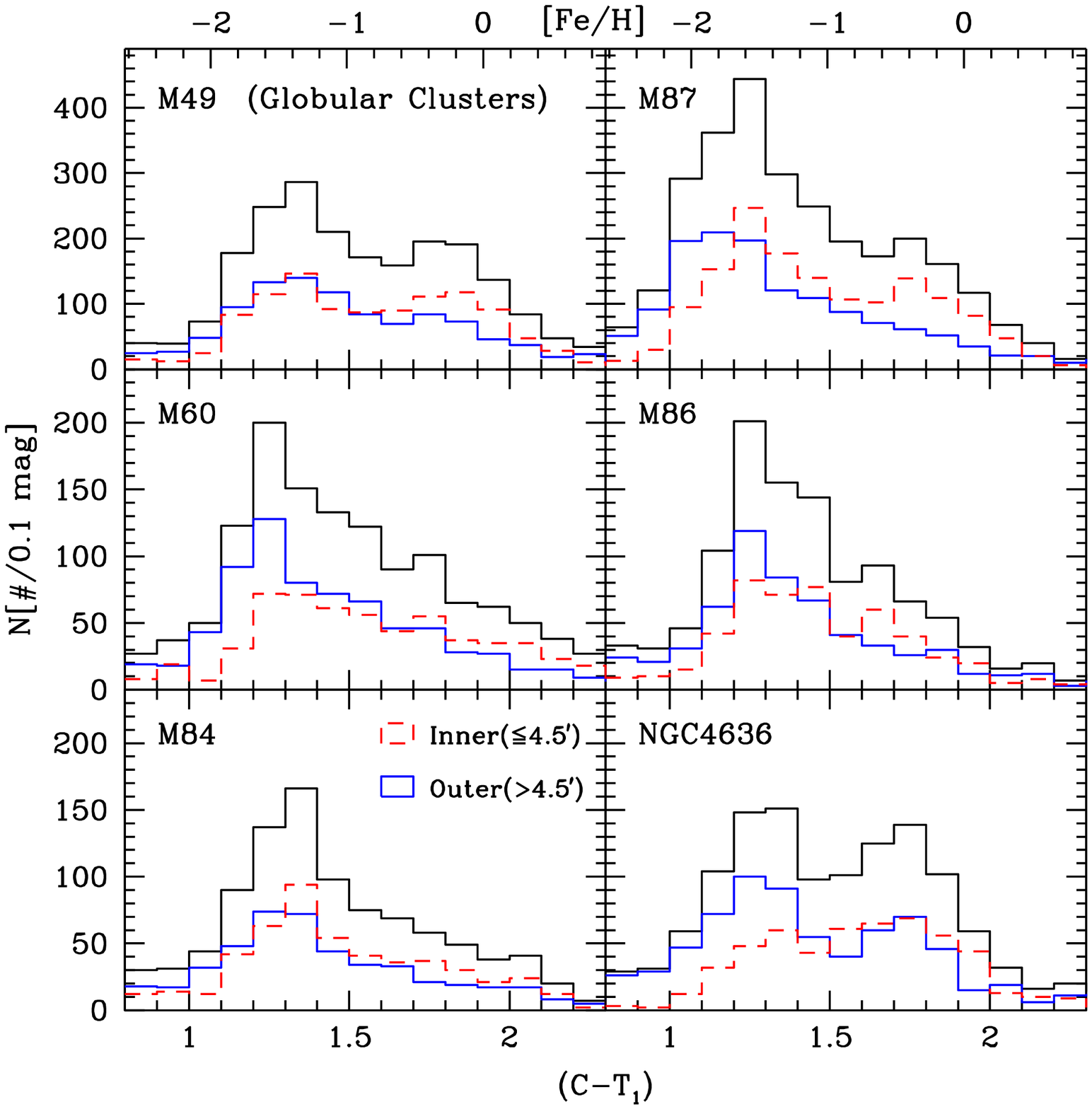} %virgocd.ps}
  \end{center}
\vskip -0.7cm
\caption{$(C-T_1)$ color distribution of bright globular cluster candidates
with $19<T_1<23$ mag in the giant elliptical galaxies in Virgo.
Objects with $1.0<(C-T_1)<2.1$ are mostly genuine globular
clusters. Red, blue and black lines represent, respectively,
 the objects in the inner region ($R<4'.5$), in the outer region ($R>4'.5$),
 and in the entire region, where $R$ is a radial distance from the center of 
 each galaxy. 
 Metallicity scale based on $(C-T_1)$ color is also shown on the top panel.
 Note the variation in the bimodality among the six galaxies that are all
 in the same Virgo cluster. (source: Lee et al. 2003)}
\label{fig6} \end{minipage}
\end{figure*}

\begin{figure*}
\begin{minipage}{\textwidth}  \begin{center}
    \leavevmode
    \epsfxsize = 16.0cm
    \epsfysize = 16.0cm
    \epsffile{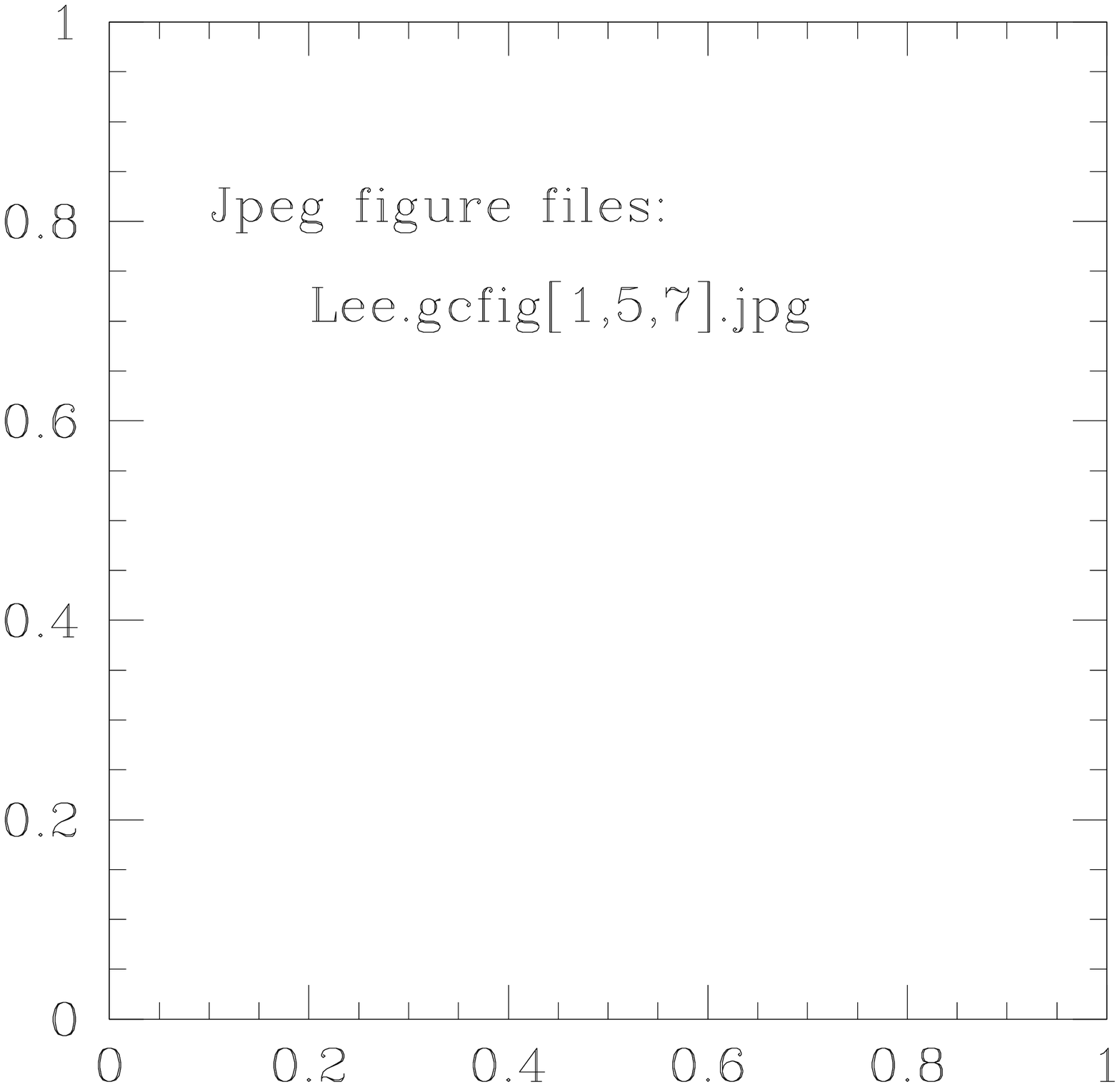} %Lee.gcfig7.ps} %m49gcsp.ps}
  \end{center}
\caption{Gray-scale map of a $T_1$ CCD images of M49 (a),
and the surface number density maps for the globular clusters with $T_1<23$ mag
for the entire GCs (b), the BGCs (c), and the RGCs (d).
Note the spatial distribution of the RGCs is similar to that of the stellar 
halo, while that of the BGCs is almost spherical.
(source: Lee, Kim, \& Geisler 1998)}
\label{fig7} \end{minipage}
\end{figure*}

Figure 6 illustrates $(C-T_1)$ color distribution of
bright globular cluster candidates
with $19<T_1<23$ mag in the same galaxies as shown in Figure 5.
Objects with $1.0<(C-T_1)<2.1$ are mostly genuine globular
clusters. Red, blue and black lines represent, respectively,
 the objects in the inner region ($R<4'.5$), in the outer region ($R>4'.5$),
 and in the entire region, where $R$ is a projected radial distance from the center of each
 galaxy. Metallicity scale based on $(C-T_1)$ color is also shown above the top panel.

 All six galaxies show clearly bimodal color distribution with peaks at
 $(C-T_1)\approx 1.3$ and $\approx 1.75$.
 %These data are being analyzed and these numbers are still preliminary.
 Interestingly, the mean metallicity difference between the
 RGCs and BGCs in all six galaxies is found to be $\Delta$[Fe/H]$\approx 1.0$ dex,
 showing the RGCs are 20 times more metal-rich than the BGCs in all galaxies.
 Why %all about 20 times
 is the difference in metallicity (all about 20 times) so similar?
Below I use the BGCs for metal-poor globular clusters and the RGCs for metal-rich
 globular clusters.

 However, the ratio of the total number between the BGCs and RGCs varies significantly among the galaxies.
 Note the dramatic contrast between M84 and NGC 4636: the BGC is much stronger than the RGC in M84,
 while both are comparable in NGC 4636.

 In addition, the ratio of the number between the BGCs and RGCs changes significantly
 depending on the galactocentric distance: relatively more RGCs are seen in the inner region than in
 the outer region.

\subsection{Spatial Structure}

M49  is an ideal target to compare the spatial distribution of globular clusters and stellar halo,
because it is significantly elongated
(the ellipticity at $r_e$(= 120 arcsec = 10.1 kpc)=0.17), Kim, Lee, \& Geisler 2000).
M87 has three times more globular clusters than M49, but its ellipticity is almost zero (E0).
To date M49 is the only giant elliptical galaxy,
for which the details of the spatial structure including the
ellipticity of the globular cluster systems were studied.
Similar studies of other giant elliptical galaxies are being done now (Lee et al. 2003).

Figure 7 displays spatial distribution of globular clusters in M49 in comparison with the stellar halo
of M49 based on the ground-based Washington photometry (Lee, Kim, \& Geisler 1998).
Figure 7 shows a $T_1$ image of M49 (a) and the surface number density maps for the entire GCs (b),
the BGCs (c), and the RGCs (d).
Note the spatial structure of the RGC system is elongated very similarly to that of the stellar halo,
while that of the BGCs is almost circular.
The RGCs are more centrally concentrated than the BGCs, which is also seen in Figure 8.

\begin{figure}[t]
  \begin{center}
    \vskip -0.7cm
    \leavevmode
    \epsfxsize = 8.7cm
    \epsfysize = 8.7cm
    \epsffile{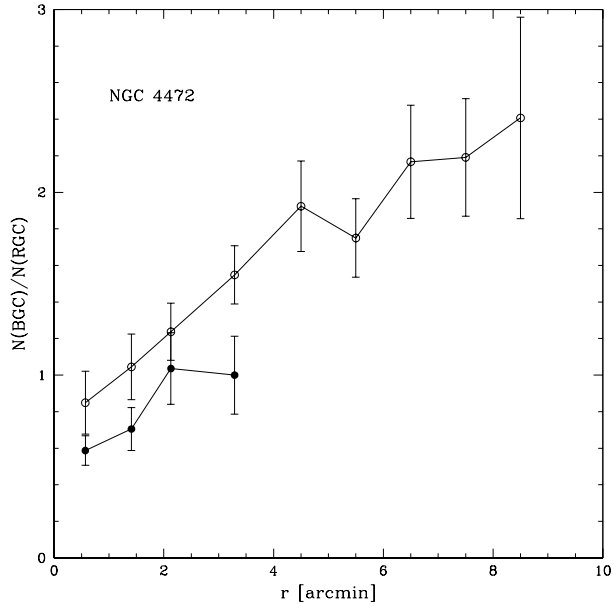} %m49ngc.ps}
  \end{center}
\vskip -0.3cm
\caption{Radial variation of the number of the BGCs to that of the RGCs with 
$V<23.9 $ mag in M49 derived from the $HST$ WFPC2 $VI$ data (filled circles).
The open circles represent the data for the globular clusters with $T_1<23$ mag
in the outer
region of M49 derived from the Washington photometry (source: Lee \& Kim 2000)}
\label{fig8}
\vskip -0.2cm
\end{figure}

Figure 8 plots the radial variation of the ratio of the number of the BGCs to that to the
RGCs. The filled circles represent the data for globular clusters with $V<23.9$ mag
based on the HST WFPC2 $VI$ photometry, and the open circles
the data for globular clusters with $T_1<23$ mag based on the ground-based Washington photometry
(Lee \& Kim 2000).
Figure 8 shows that the N(BGC)/N(RGC) increases as the galactocentric radius increases.
There are about equal number of BGCs and RGCs in the inner region, but the outer region is dominated
by the BGCs.

This shows the spatial structure of the RGC system is very consistent with that of the stellar halo, while
that of the BGC system is different from both.
So it is the RGCs, not the BGCs, that go with halo stars in M49.

\subsection{Color Gradients}

\begin{figure*}
 \begin{minipage}{\textwidth}
 \vspace{2.8in}
 \begin{center}
    \leavevmode
    \epsfxsize = 20.0cm
    \epsfysize = 20.0cm
    \epsffile{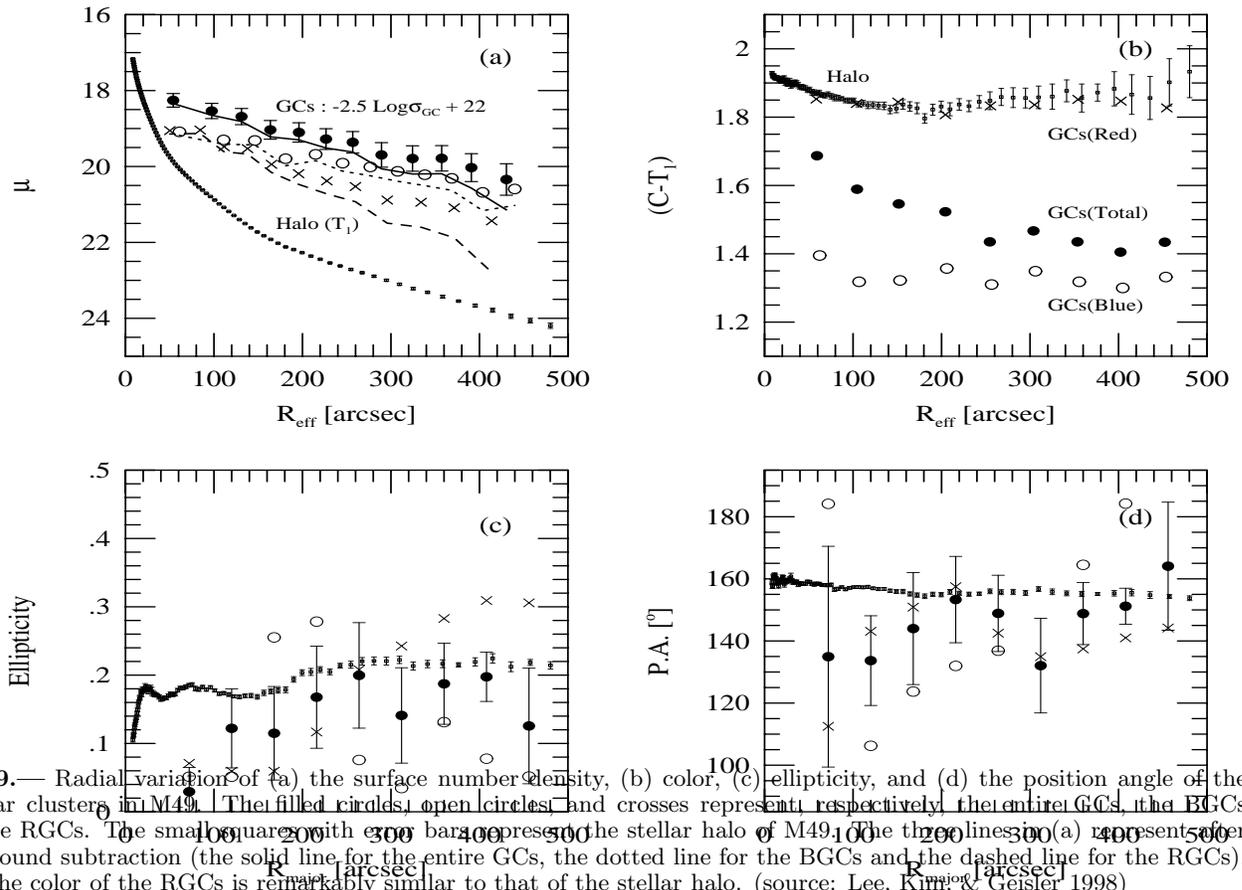} %m49gchalo.ps}
  \end{center}
  \vspace{-3.8in}
\caption{Radial variation of (a) the surface number density, (b) color, 
(c) ellipticity, and (d) the position angle of the globular clusters in M49. 
The filled circles, open circles, and crosses represent, respectively, the entire GCs,
the BGCs and the RGCs. The small squares with error bars represent the stellar 
halo of M49.
The three lines in (a) represent after background subtraction (the solid line 
for the entire GCs,
the dotted line for the BGCs and the dashed line for the RGCs).
Note the color of the RGCs is remarkably similar to that of the stellar halo.
(source: Lee, Kim, \& Geisler 1998)}
\label{fig9} \end{minipage}
\end{figure*}

Figures 9 and 10 display color gradients of the globular clusters in M49 in comparison with the stellar halo
(Lee, Kim, \& Geisler 1998, Lee \& Kim 2000, Kim, Lee, \& Geisler 2000).
Striking features seen in these figures are:
1) all GCs show a strong radial color gradient (much stronger than the stellar halo);
2) the color profile of the RGCs is almost the same as that of the stellar halo
(note, however, that the number density profile of the RGCs is much more extended than that
of the stellar halo, as seen in Figure 9(a)); and
3) the BGCs and RGCs show, respectively, little radial color gradients.
The strong color gradient of all GCs is primarily due to the stronger central concentration of the
RGCs compared with the BGCs.
This evidence shows again that the RGCs and the halo stars in M49 are of the same origin.

\subsection{Luminosity Function}

\begin{figure}[t]
  \begin{center}
    \vskip -1.4cm
    \leavevmode
    \epsfxsize = 8.8cm
    \epsfysize = 8.8cm
    \epsffile{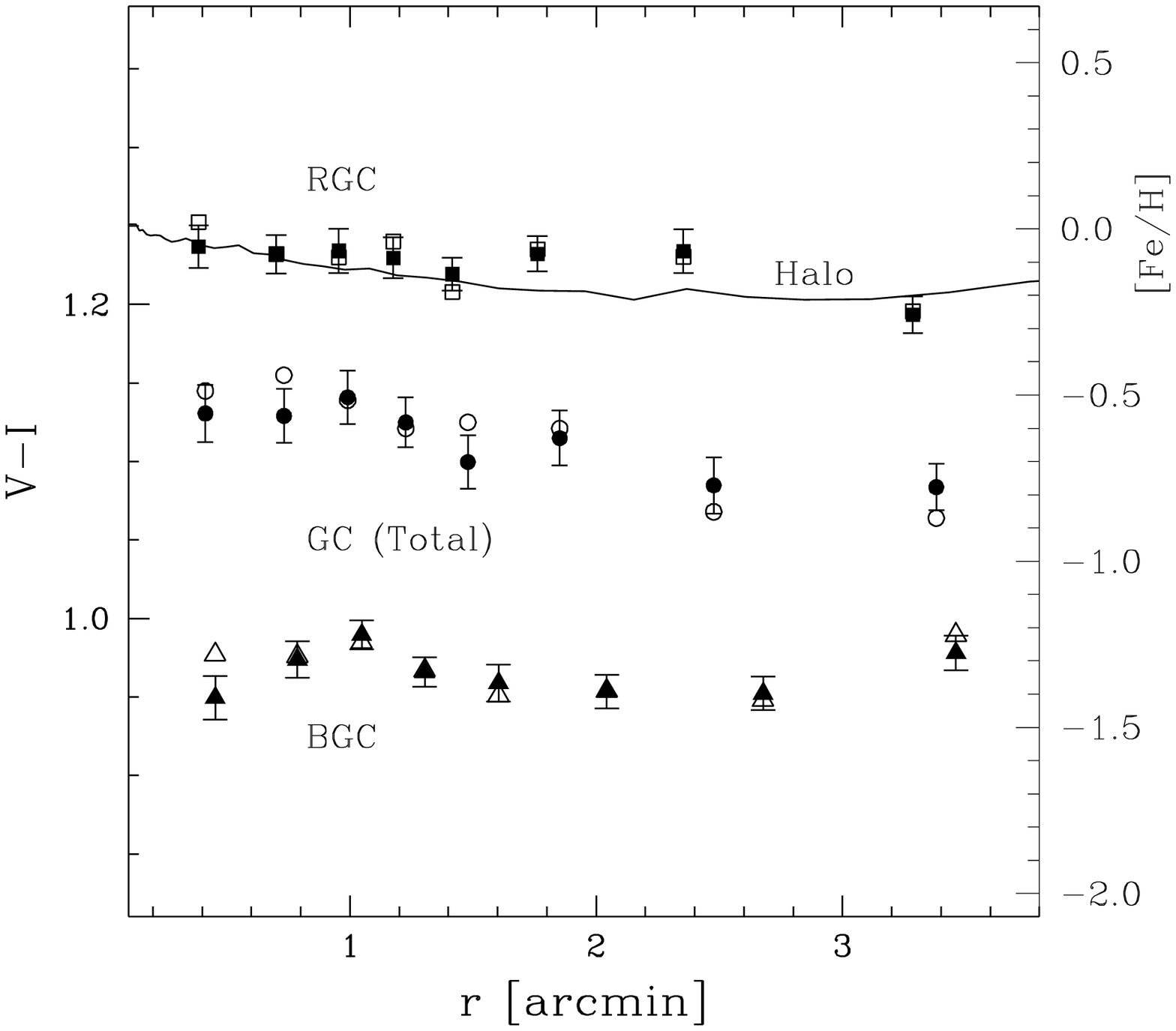} %m49gcvi.ps}
  \end{center}
\vskip -1.0cm
\caption{Radial variation of the mean (closed symbols) and median (open symbols)
color of the bright globular clusters with $V<23.9$ mag in M49.
The circles, triangles, and squares represent, respectively, the total sample, 
the BGCs and the RGCs.
The solid line represents the mean color of the stellar halo
 of M49 given by Kim, Lee, \& Geisler (2000).
 Note the color of the RGCs is remarkably similar to that of the stellar halo.
 (source: Lee \& Kim 2000)}
\label{fig10}
\vskip -0.3cm
\end{figure}

Luminosity functions of old globular clusters in elliptical galaxies are approximately fit by
a Gaussian with a peak at $M_V=-7.4$ mag and a width $\sigma\approx 1.2$
(or by a $t_5$ function),
which is often used as distance indicator.
Figure 11 illustrates the luminosity functions of globular clusters in M49 based on the HST WFPC2 $VI$
photometry, showing a peak at $V=23.50\pm0.16$ and $I=22.40\pm0.14$, leading to a distance estimate
of $d=14.7\pm1.3$ Mpc (Lee\& Kim 2000).
The bright part of the luminosity function of globular clusters in elliptical galaxies is approximately
fit by a power law, and the luminosity functions of bright blue clusters in the merging galaxies are also
well fit by a power law. The difference in the faint part between these two kinds
is considered to be due to dynamical evolution on the low-mass clusters,
which converts the power-law luminosity functions into a Gaussian form
or two component power-law form (Fall \& Zhang 2001, Smith \& Burkert 2002).
The bright part of the luminosity function is less affected than the faint part
by the dynamical evolution of clusters.

One thorny problem in the luminosity function is the difference in the peak magnitude between the BGC and RGC.
The peak magnitude of the globular cluster luminosity function may depend on metallicity,
age, and dynamical evolution.
In the case of M49, Puzia et al. (1999) found, using the same HST WFPC2 $VI$ images for three fields in M49
as used by Lee \& Kim (2000),
that the peak magnitude of the RGC is fainter than that of the BGC, by $\Delta V(RGC-BGC)=0.51$ mag
and $\Delta I=0.41$ mag, and concluded, from this,
that the BGCs and the RGCs are coeval within the errors
of $\sim 3$ Gyr. Later Larsen et al. (2001), using the same HST data,
concluded that the  peak magnitude of the RGC is fainter than that of the BGC, by
$\Delta V=(24.21\pm0.23)-(23.38\pm0.16)=0.83$ mag, and the peak magnitude for the entire sample is
$V=23.78^{+0.14}_{-0.15}$, which is 0.38 mag fainter than the value given by Lee \& Kim (2000).

On the other hand, Lee \& Kim (2000) found, using the HST WFPC2 $VI$ images for four fields,
that the $V$-band peak magnitudes of the RGC and the BGC are similar within the errors:
$V(BGC)=23.53\pm0.16$, $V(RGC)=23.44\pm0.22$, $I(BGC)=22.63\pm0.17$ and $I(RGC)=22.30\pm0.26$ mag,
as shown in Figure 11. In this case, the $I$-band magnitude difference, $\Delta I = -0.33$ is consistent with
the $(V-I)$ difference between the RGC and BGC, $\Delta(V-I)= 1.233-0.975=0.26$
(i.e., the $I$-band peak magnitude of the RGC is brighter than that of the BGC, if the $V$-band
peak magnitudes of the BGC and RGC are the same.). This result is also consistent with previous
results based on the ground-based Washington photometry given by Lee, Kim, \& Geisler (1998) and the similar
result for M87 by Kundu et al. (1999).
This result indicates that the RGCs may be several Gyr younger than the BGCs, if the same isochrones
as used by Puzia et al. (1999) are used.

It is puzzling that the two groups's results based on the basically same data gave significantly
different estimates for the peak magnitudes of the BGC and RGC in M49.
The reason for this difference appears to be, in part, due to the difference in the method of photometry.
Comparison of the luminosity functions for the same fields given by two groups shows that
Lee \& Kim (2000)'s photometry is more complete in the faint end than Puzia et al.(1999)'s
even before incompleteness correction (see Figure 9 in Lee \& Kim 2000).
This indicates that the careful analysis is needed for investigating the difference
in the peak magnitude between the BGCs and RGCs in distant galaxies.

\subsection{Specific Frequency}

\begin{figure*}
 \begin{minipage}{\textwidth}
 \vspace{1in}
 \begin{center}
    \leavevmode
    \epsfxsize = 16.0cm
    \epsfysize = 16.0cm
    \epsffile{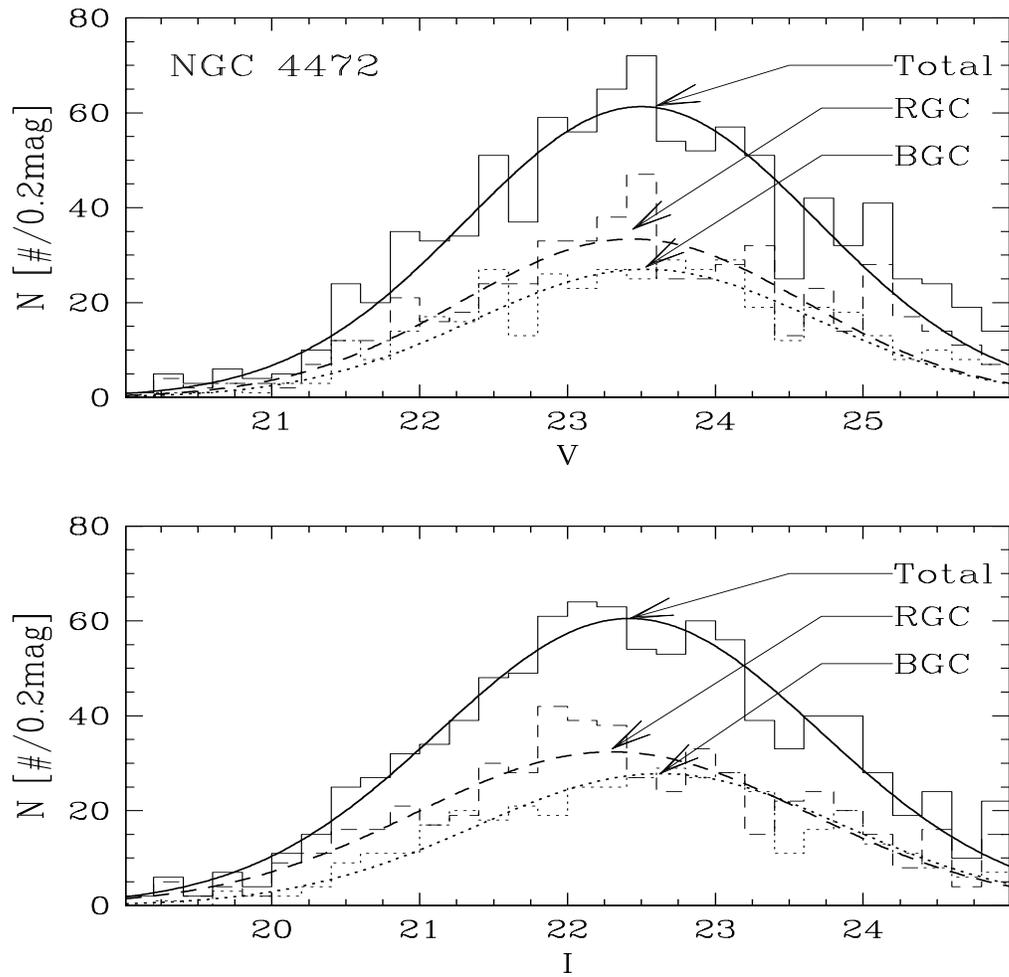} %m49gclf.ps}
  \end{center}
  \vspace{-0.8in}
\caption{$V$-band and $I$-band luminosity functions of the globular clusters 
in M49 derived from
$HST$ WFPC2 images of four fields (histograms) fit with
the Gaussian functions (the curved lines). (source: Lee \& Kim 2000)}
\label{fig11} \end{minipage}
\end{figure*}

The globular cluster specific frequency is defined as the total number of globular clusters
per unit host galaxy luminosity: $S_N=N(GC) 10^{0.4(M_V +15)}$
(Harris \& van den Bergh 1981).
$S_N$ for dwarf and giant elliptical galaxies in rich clusters
($S_N \approx 4-6$)
is higher than $S_N$ for such galaxies in loose groups ($S_N \approx 2-3$),
and $S_N$ for spiral galaxies is as low as about 1 (Harris 1991).
Several brightest cluster galaxies and cD galaxies show very high $S_N = 10 - 20$ (see
Figure 8 in Beasley et al. 2002).
The higher $S_N$ for ellipticals compared with $S_N$ for spiral galaxies motivated a merger model
(Ashman \& Zepf 1992).

Recently C\^ote et al. (2000) pointed out that, if the bulge luminosity is used instead of total galaxy
luminosity, $S_N$ for 11 spiral galaxies (Sa to Sc) in the sample of Kissler-Patig et al. (1999)
 becomes $S_N=3.8\pm2.9$, which is comparable to that for elliptical galaxies.
The high $S_N$ ($>5$) for giant elliptical galaxies in rich clusters are primarily due to the BGCs, rather than
due to RGCs (Forbes et al. 1997).
Also the local $S_N$ increases as the galactocentric distance increases, which is due to the fact that
the BGCs are relatively more abundant in the outer region than the RGCs (Lee, Kim, \& Geisler 1998).

\subsection{Kinematics}

\begin{figure*}
\begin{minipage}{\textwidth}
  \begin{center}
    \leavevmode
    \epsfxsize = 16.0cm
    \epsfysize = 16.0cm
    \epsffile{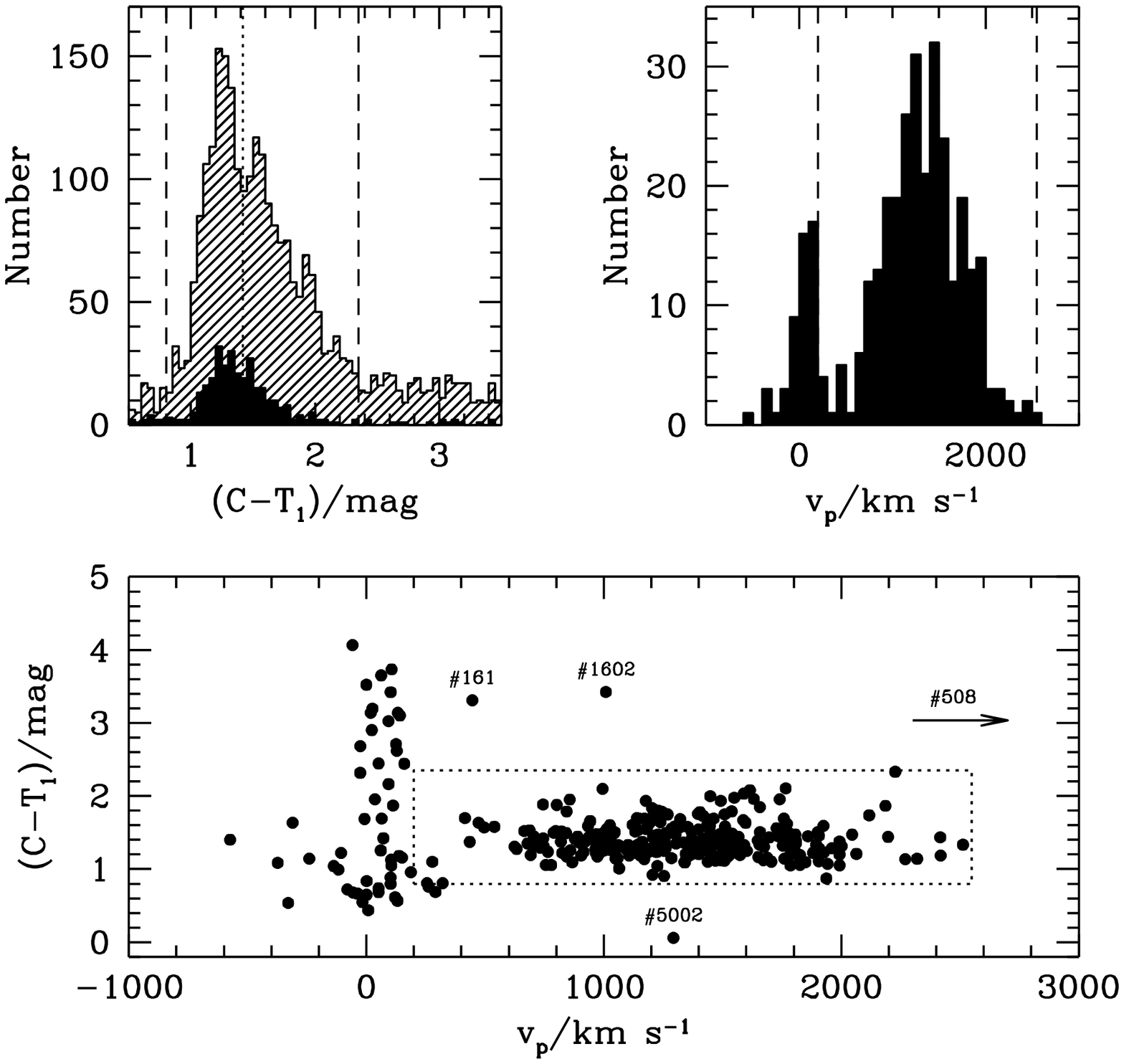} %cote01m87v.ps}
  \end{center}
  \vspace{-0.1in}
\caption{(Upper left panel) Color distribution of globular cluster candidates in
 M87
with 334  measured radial velocity (black histogram) of C\^ote et al. (2001)
 among the photometric sample (shaded histogram).
(Upper right panel) Radial velocity distribution of globular cluster candidates 
in M87.
(Lower panel) $(C-T_1)$ vs. radial velocities of the globular cluster candidates
 in M87.
There are 278 globular clusters inside the dashed line. 
(source: C\^ote et al. 2001)}
\label{fig12} \end{minipage}
\end{figure*}

Figure 12 displays colors and radial velocities of globular cluster candidates in M87:
a) the color distribution of globular cluster candidates in M87
with 334  measured radial velocity (black histogram) of C\^ote et al. (2001)
 among the photometric sample (shaded histogram),
b) the radial velocity distribution of globular cluster candidates, and
c) $(C-T_1)$ vs. radial velocities of the globular clusters in M87.
Figure 12 shows 1) 278 objects inside the dashed box (with $0.8<(C-T_1)<2.35$) are genuine globular clusters
belonging to M87 (or at least to the Virgo cluster),
2) most objects with colors $(C-T_1)>2.35$ or $(C-T_1)<0.8$ are foreground stars, 
and 3)
the fraction of objects with colors $0.8<(C-T_1)<2.35$ is negligible compared with the globular clusters
with the same color range.
This confirms that most of the photometric globular cluster candidates based on the Washington photometry
are genuine globular clusters, showing the high efficiency of the Washington photometry in selecting globular
cluster candidates.

\begin{table*}[t]
\begin{minipage}{\textwidth}
\vspace{0.4cm}
\begin{center}

{\bf Table 1.}~~~Kinematics of Globular Clusters in gEs
\begin{tabular} {c c c c c c l l}
\hline\hline
Galaxy & GC & $<\sigma_p>$ & $<\Omega R >$ & $<\Omega R / \sigma_p>$ & N(GC) & Ref.  \\
\hline
M87 & All &  $383^{+31}_{-7}$   & $171^{+39}_{-30}$  & $0.45^{+0.09}_{-0.09}$  & 278 & 1 \\
    & BGC &  $397^{+36}_{-14}$  & $186^{+58}_{-41}$  & $0.47^{+0.13}_{-0.11}$  &     & 1 \\
    & RGC &  $365^{+38}_{-18}$  & $155^{+53}_{-37}$  & $0.43^{+0.14}_{-0.12}$  &     & 1 \\
\hline
M49 & All &  $316^{+27}_{-8}$  & $48^{+52}_{-26}$  & $0.15^{+0.15}_{-0.08}$  & 263 & 2 \\
    & BGC &  $342^{+33}_{-18}$  & $93^{+69}_{-37}$  & $0.27^{+0.19}_{-0.11}$  &    & 2 \\
    & RGC &  $265^{+34}_{-13}$  & $-26^{+64}_{-79}$  & $0.10^{+0.27}_{-0.25}$  &   & 2 \\
\hline
NGC 1399 & All & $304\pm11$   & $ $  &  & 470 & 3,4 \\
         & BGC & $297\pm16$   & $ $  &  & & 3,4 \\
         & RGC & $355\pm22$   & $ $  &  & & 3,4 \\
\hline
MWG & BGC &   &  & 0.32  & & 5 \\
 & RGC &   &  & 1.05  & & 5 \\
\hline
M31 & BGC &   &  & 0.85  & & 6 \\
 & RGC &   &  & 1.10  & & 6 \\
    \hline
\end{tabular}
\end{center}
References: (1) C\^ote et al. (2001); (2)C\^ote et al. (2003); (3) Richtler et al. (2002);
(4) Dirsch et al. (2002); (5) C\^ote (1999); (6) Perrett et al. (2002).

\end{minipage}
\vspace{-0.3cm}
\end{table*}

There are only three giant elliptical galaxies for which the kinematics of globular clusters were
studied:
M87, M49 and NGC 1399 (Cohen \& Ryzhov 1997, Kissler-Patig \& Gebhardt 1998, Kissler-Patig et al. 1998,
Minniti et al. 1998, Zepf et al. 2000, C\^ote et al. 2001, 2003).
C\^ote (2002) gave a nice summary of kinematics of globular clusters in  M49 and M87.
Table 1 is a slightly updated version of C\^ote (2002)'s compilation, including NGC 1399.
Table 1 lists the projected velocity dispersion, $<\sigma_p>$,
the rotational  velocity, $<\Omega R >$, and
the anisotropy parameter $<\Omega R / \sigma_p>$.

What is surprising is that all three galaxies showed different kinematics of globular clusters:
1) M87 GCs (in total) show a larger velocity dispersion than M49 GCs, although M87 is somewhat fainter than M49.
This must be related with the fact that M87 is located right at the center of the Virgo cluster, while
M47 is far from the center of the Virgo;
2) The BGCs have a larger velocity dispersion than the RGCs in M49, but NGC 1399 GCs show an opposite
result. On the other hand, M87 GCs show a similar velocity dispersion between the BGC and the RGC;
3) The M87 GCs rotate faster than the M49 GCs,  and there is little difference in the rotation velocity between
the BGC and RGCs in M87. In contrast, there is significant difference in the rotation velocity between
the BGC and RGCs in M49. The BGCs in M49 rotate with a velocity of 93 km s$^{-1}$, while the RGCs in M49
show little rotation. The kinematics of the M49 GCs is consistent with the gaseous merger model (Ashman \& Zepf 1992),
but that of the M87 GCs is not.
4) The BGCs in M87, M49, and our Galaxy show similar anisotropy ($<\Omega R / \sigma_p>$),
but smaller than that of the BGCs in M31.
However, the RGCs in M87 and M49 show much smaller anisotropy  to the RGCs of in our Galaxy and M31.
One notable feature is that  the BGCs in M31 is rotating almost as fast as the RGCs.
In summary, any uniform feature is not yet emerging from the kinematics of giant elliptical galaxies.

\subsection{Age}

\begin{table*}[t]
\begin{minipage}{\textwidth}
\vspace{0.4cm}
\begin{center}

{\bf Table 2.}~~~ Age estimates for GCs in gEs
\begin{tabular} {c c c c c c}
\hline\hline
Galaxy & t(BGC) & t(RGC) & $\Delta t$ & Source & Ref.  \\
\hline
M87 & $13.7\pm1.8$ & $12.7\pm2.2$ & $1.0\pm3.3$   & Spectra  & 1 \\
    &              &              & $\approx 3-6$ & LF($VI$)  & 2 \\
    &              &              & $0.2\pm2.0$   & LF($uvby$)  & 3 \\
    & $>15$        & $>15$        &               & Sp(+TMB02) & 4 \\
    &              &              & --a few       &$(V-I)$-$(V-K)$ & 5 \\
\hline
M49 &              &              & $-0.6\pm3.2$  & LF($VI$)  & 6 \\
    &              &              & $\approx 3-6$ & LF($VI$)  & 7 \\
    & $14.5\pm4$   & $13.8\pm6$   & $0.7\pm7.2$   & Spectra       & 8 \\
    & $13.9\pm4.1$ & $12.8\pm2.6$ &               & Spectra (+W94) & 4 \\
    & $>15\pm2$ & $(5-10)\pm(3-4)$ &                & Sp (+TMB02) & 4 \\
\hline
NGC 1399 & 11(N=8)   & 2? (N=2)    &            & Spectra & 9 \\
\hline
NGC 3115 &          &              & $-2\pm3$ & $(V-I)$-$(V-K)$ & 10 \\
   \hline
\end{tabular}
\end{center}
References: (1) Cohen, Blakeslee, \& Ryzhov (1998); (2) Kundu et al. (1999); (3)
 Jord\'an et al. (2002a);
(4) Cohen, Blakeslee, \& C\^ote (2003) used two models: Worthey (1994, W94) and
Thomas, Maraston, \& Bender (2003, TMB03);
(5) Kissler-Patig, Brodie, \& Minniti (2002); (6) Puzia et al. (1999); (7) Lee \& Kim (2000);
(8) Beasley et al. (2000);
(9) Forbes, D. A.  et al. (2001) (the ages of the two metal-rich globular clusters
may be 2 Gyr or $>15$ Gyr);
(10) Puzia et al. (2002) who gave a mean of estimates based on four different models.
\end{minipage}
\vspace{-0.3cm}
\end{table*}

Age determination of globular clusters in giant elliptical galaxies is extremely
difficult, and currest age estimates include a large uncertainty.
There are only a few giant elliptical galaxies for which the ages of the globular clusters were estimated.
Table 2 is an updated version of C\^ote (2002)'s compilation of the age estimates of M87 and M49,
by including other references. %Lee \& Kim (2000), Cohen, Blakeslee, \ as described above.
Table 2 shows 1) the globular clusters in these galaxies are older than 10 Gyr,
and 2) the errors of age estimates of the globular clusters are large.
The BGCs and RGCs may be said to be coeval within the error of 3-7 Gyr.
However, this uncertainty is too large to distinguish different models for the formation of
globular clusters, because the age difference
between the BGC and the RGC is expected to be smaller than 3 Gyr (see the final section).

The sources of uncertainties are both observational data and the model isochrones
(Thomas, Maraston, \& Bender 2003, Thomas \& Maraston 2003).
Cohen, Blakeslee, \&  C\^ote (2003) derived, from the comparison of their spectra for 47 globular
clusters in M49 with the models of Worthey (1994) and Thomas, Maraston, \& Bender (2003),
ages which are significantly different between the two models, as listed in Table 2.
They pointed out ``However, the only safe statement in our view is that M49 GCs are in the mean older than
10 Gyr''.
Puzia et al (2002) also found a large difference in age estimates depending on isochrone models, when
they estimates the age of globular clusters  in NGC 3115 using $(V-I)$--$(V-K)$ diagrams
(note that this method was used also for estimates of age and metallicity of early-type galaxies
in galaxy clusters (Smail et al. 2001)).
Better and more age estimates are needed.

\subsection{Relation with Host Galaxies}

\begin{figure*}
    \vspace{0.3in}
    \begin{minipage}{\textwidth}  \begin{center}
    \leavevmode
    \epsfxsize = 16.0cm
    \epsfysize = 16.0cm
    \epsffile{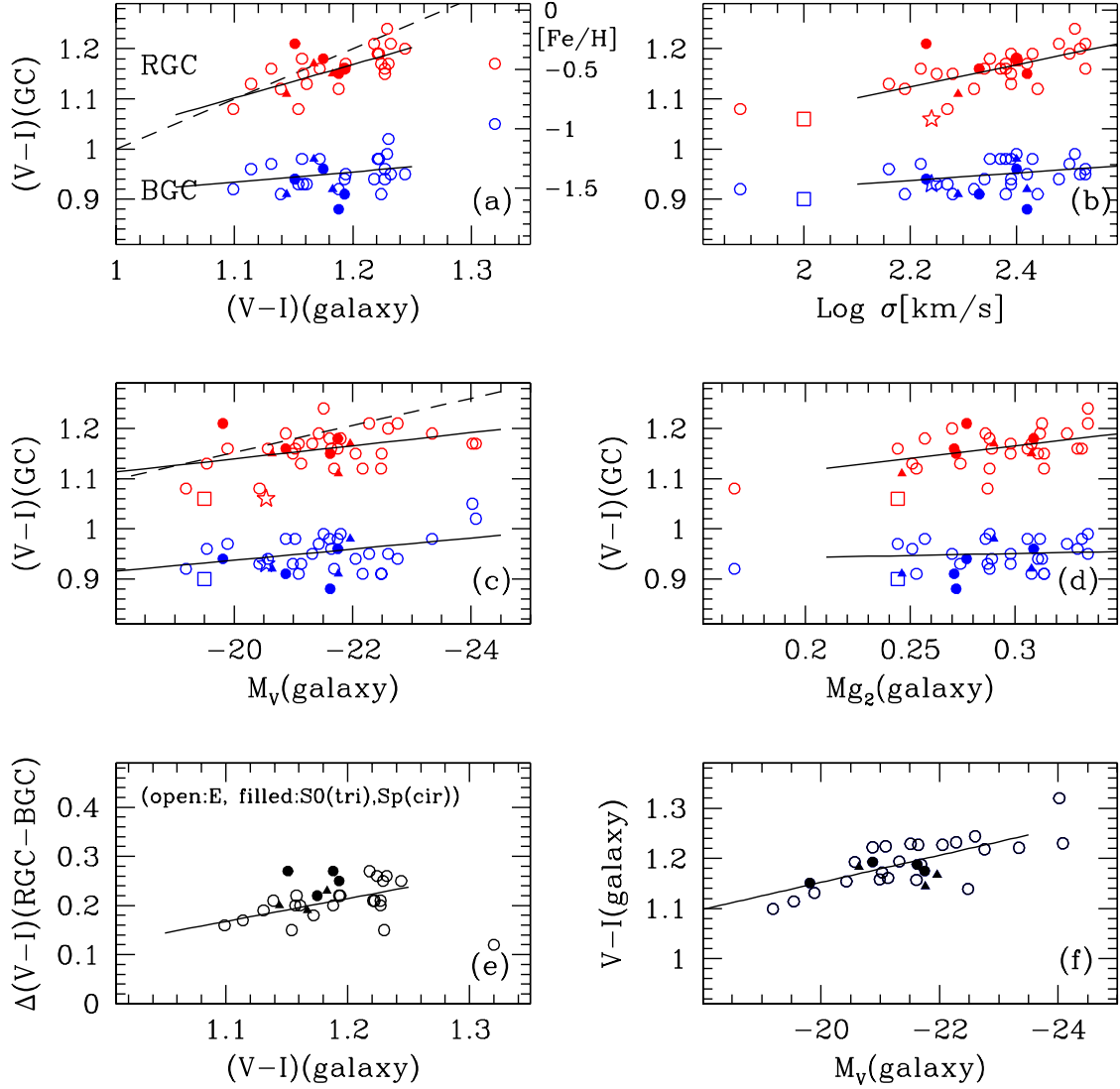} %egcall.ps}
  \end{center}
  \vspace{-0.3in}
\caption{Relations between the globular clusters and their host
galaxies with early types.
(a) $(V-I)$(GC) vs. $(V-I)$(galaxy).
(b) $(V-I)$(GC) vs. Log $\sigma_v$(galaxy), velocity dispersion.
(c) $(V-I)$(GC) vs. $M_V$(galaxy).
(d) $(V-I)$(GC) vs. $Mg_2$(galaxy).
(e) $\Delta (V-I)$(RGC-BGC) vs. $(V-I)$(galaxy).
(f) $(V-I)$(galaxy) vs. $M_V$(galaxy).
The open circles, filled circles, filled triangles represent,
respectively, Es, S0s and galaxies with types later than S0.
Open square and star symbol represent, respectively, our Galaxy and M31
for which the $(V-I)$ colors were predicted from [Fe/H].
Solid lines represent the linear fits to E's with $HST$ WFPC2 photometry.
The dashed line in (a) represent the one-to-one
relation, and the dashed line in (c) represents the color-magnitude relation of
host galaxies which is also plotted in (f) as a solid line.
Sources of the data: Barmby et al. (2000),
Larsen et al. (2001), Forbes \& Forte (2001), Kundu \&
Whitmore (2001a,b), Lee et al. (2002), Golev \& Prugniel (1998),
and Tonry et al. (2001).
}
\label{fig13} \end{minipage}
\end{figure*}

The correlation between the RGCs and their host galaxies has been pointed out by Forbes, Brodie, \&
Grillmair (1997), Forbes \& Forte (2001) and Larsen et al. (2001), although Kundu \& Whitmore (2001a) argued that
the correlation is weak at best ,or disappears when the galaxies with strongest bimodality are considered.
Figure 13 displays the colors of the globular clusters in elliptical galaxies versus the parameters for their host galaxies.
Our Galaxy and M31 are included for comparison.
The sources of the data are:  Forbes \& Forte (2001), Kundu \& Whitmore (2001a,b), Larsen et al. (2001),
Lee \& Kim (2000), Lee et al. (2002) for globular clusters,
Tonry et al. (2001) and Jensen et al. (2003) for distances of the galaxies,
Golev \& Prugniel (1998) for $Mg_2$ of the galaxies, and
Barmby et al. (2000) for the globular clusters in M31 and our Galaxy.
There are 42 early-type galaxies (except for our Galaxy and M31) in the sample, and
22 to 28 elliptical galaxies with $HST$ observations among them were used
for linear fitting of the data (represented by the solid line).

Figure 13 shows several important features:
1) The $(V-I)$ colors of the RGCs are very similar to those of the host galaxies.
2) The $(V-I)$ colors of the RGCs show a correlation with those of their host galaxies,
while those of the BGCs do little.
3) The slope of the $(V-I)$ colors of the RGCs vs. the color of their host galaxies
(the solid line in Figure 13(c)) is somewhat flatter than that for the color-magnitude relation of the host galaxies
(the dashed line in Figures 13(a) and the solid line in Figure 13(f)).
4) The difference in color between the BGCs and the RGCs is shown in Figure 13(e).
The color difference is  in the range of $0.15<\Delta (V-I)<0.27$, which corresponds to the
metallicity range of $0.63<\Delta [{\rm Fe/H}]<1.14$ dex, using
[Fe/H]$=4.22(\pm0.39)(V-I)_0 - 5.39(\pm0.35)$ derived from M31 globular clusters (Barmby et al. 2000).
5) Similar trends are seen against the logarithmic value of velocity dispersion, the total magnitude,
and the $Mg_2$ of the host galaxies (Figure 13(b), 13(c), and 13(d)).
These results show that the RGCs show a close relation with the stellar halos of their host galaxies, while
the BGCs do not.

\subsection{Young to Intermediate-Age Globular Clusters in Interacting/Merging Galaxies}

Star clusters brighter and younger than typical globular clusters in our Galaxy
have been found in several merging galaxies with young to intermediate-age and
in a few dynamically young elliptical galaxies (such as NGC 3610 and NGC 1700)
(Whitmore 2000, Schweizer 2002b, Davies et al. 2001, Larsen et al. 2003). 
They appear to follow well the age sequence
in the $\Delta V_{10}-\Delta (V-I)$ diagram where $\Delta V_{10}$ is a magnitude
difference between the 10th-bright 2nd-generation GC and its old counterpart
(see Figure 6 in Schweizer 2002b).
This evidence shows clearly that massive star clusters formed during the merger of
galaxies, and the mering of galaxies happen continuously until recently.

\subsection{Summary}

Several observational evidences show that the RGCs in elliptical galaxies
are closely related with the stellar halo
of their host galaxies in color and spatial distribution, while the BGCs are not.
Three giant elliptical galaxies show all different kinematics of the globular clusters.
Present age estimates of the BGCs and the RGCs contain large uncertainty so that the difference
in age between the two are not yet known, but show that the BGCs and the RGCs may
be both older than 10 Gyr (see also Ashman \& Zepf 2001).

\section{MODELS OF FORMATION OF GLOBULAR CLUSTERS}

Models for the formation of globular clusters are often classified
according to the formation epoch of globular clusters:
Primary, secondary and tertiary formation models, or
pre-galactic, proto-galactic, and post-galactic models (see C\^ote 2002).
Here I divide the models according to the models of elliptical galaxy formation (MCM and HMM):
the primary origin models (only the BGC),
the Monolithic Collapse Globular Cluster Formation Models (MCGCFM), and
the Hierarchical Merging Globular Cluster Formation Models (HMGCFM).

In the primary origin models, the BGCs formed before galaxies formed, and the RGCs
are not mentioned.
In the MCGCFM, the BGCs and then the RGCs formed during the multiple collapses
of a large proto-galactic cloud.
In the HMGCFM,
the BGCs and RGCs formed in the regime of the HMM.

\subsection{Primary Origin Models:before Galaxies}

Peebles \& Dicke (1968) proposed, noting the little variation of the luminosities of globular clusters,
that globular clusters formed from the gravitationally bound gas clouds before galaxies formed
soon after the recombination era, when the temperature goes down about 4000 K and the density
is about $10^{-4}$ atoms cm$^{-3}$.
Right after the recombination era, the Jeans radius is about 5 pc, and the
Jeans  mass  $10^5-10^6$ $M_\odot$, which is similar to the mean mass of the globular clusters in
the halo of our Galaxy.
Newer versions of this model were given by Peebles (1984) and Cen(2001).
These globular clusters are expected to be very old, metal-poor and everywhere.

\subsection{The MCGCFM}

Fall \& Rees (1985) proposed that globular clusters would form in the collapsing gas of
a protogalaxy. Due to the thermal instability  in hot gaseous halos of young galaxies the cold
components are compressed into discrete clouds with temperatures near $10^4$ K,  mean
densities with 1--10 $M_\odot$ pc$^{-3}$, and masses of $\sim 10^6$ $M_\odot$.

Harris \& Pudritz (1994) noted that the mass functions of massive globular clusters ($>10^5$ $M_\odot$)
in giant elliptical galaxies and disk galaxies
have a power law index similar to that of the giant molecular clouds in the Milky Way
($N \sim M^{-1.7}$), and proposed
that globular clusters formed out of dense cores in the (hypothetical)
supergiant molecular clouds (with masses of $10^9$ $M_\odot$ and diameters of $\sim 1$ kpc) which existed
in the protogalactic epoch. The masses of these clouds are similar to those of dwarf galaxies.
They predicted that the epoch of major globular cluster formation is at $z<6$.
Numerical simulations based this concept were given by Weil \& Pudritz (2001).

Forbes, Brodie, \& Grillmair (1997) proposed an episodic in situ formation plus tidal stripping model.
In this model, metal-poor globular clusters are formed first at an early stage in the initial collapse of the protogalactic
cloud with only minor star formation. After some time (about 4 Gyr) of quiescence,
metal-rich globular clusters are
formed out of more enriched gas, roughly contemporaneously with most of the galaxy stars, during
the major collapse of the protogalactic cloud, which then forms an elliptical galaxy with two
globular cluster populations. Most of the globular clusters are formed in the first formation episode and
should not be structurally related to the halo light, while the newly formed GCs are closely
coupled to the galaxy and share a common chemical enrichment history and structural characteristics.
%prediction for kinematics-Forbes, Brodie, \& Grillmair (1997)

\subsection{The HMGCFM} %Multi Collapse Models:with Galaxies}

There are two types of models in the HMGCF. One concentrates on the major
gaseous merger of two spiral galaxies at late stage of evolution,
and the other includes both early formation
of globular clusters and late merging/accretion processes of smaller objects  of a dwarf
galaxy size (proto-galactic fragments).

\noindent {\bf (Major Gaseous Merger Models)}

Ashman \& Zepf (1992) proposed a gaseous merger model (as also Zepf \& Ashman (1993), Zepf et al. (2000)).
In the gaseous merger model, an elliptical galaxy is formed by the merging of two or more
gas-rich spiral galaxies. The spiral galaxies have enriched gas in the disk
and metal-poor GCs in the halo (which need to be explained separately). New GCs are formed
from the enriched gas during the merger/interaction process. The resulting elliptical
galaxy then has two GC populations: the younger, more metal-rich, and spatially more
concentrated clusters formed as a result of the merger, and the original, metal-poor,
spatially more extended GCs formed in the progentor spirals.

A key idea in this model was
to explain the fact that the specific frequencies in elliptical galaxies are a factor of two higher
than those in spiral galaxies. In this model, the BGCs are expected to show some rotation, while
the RGCs are expected to show little rotation, because the angular moment would be
transported to the outer region during the merging process (Zepf et al. 2000).

Bekki et al. (2002) presented an extensive simulation of globular clusters formed in merging
and interacting galaxies, assuming that the GCs formed at high redshift ($z>5$) in protoglactic
fragments, and during the subsequent gas-rich merging of these fragments.
They found that the BGCs should be $\approx
3$ Gyr older for gEs in clusters (5 Gyr for low-luminosity elliptical galaxies in the field or groups)
than the RGCs, and that the age dispersion of the RGCs  is large ($t \approx 5-12$ Gyr).
They found also that the dissipative merging of present-day spirals would produce
super-solar metal-rich globular clusters which are not observed now.
To avoid this problem, they pointed out that, if elliptical galaxies form by dissipative major merger,
then they must do it at a very early epoch.

\noindent{\bf (Proto-galactic Fragment Models)}

%Accretion/merging
C\^ote, Marzke, \& West (1998) proposed a model showing a change of paradigm:
the metal-rich globular clusters are the intrinsic population of a galaxy, and the galaxy
acquires later the metal-poor globular clusters via mergers or tidal stripping.
Later C\^ote et al. (2000) suggested that the Galactic halo and its globular cluster system
were assembled via the accretion and disruption of $\sim 10^3$ metal-poor, proto-Galactic fragments
by a proto-bulge which owns metal-rich globular cluster system (see also C\^ote, West, \& Marzke 2002).
In this model no globular clusters formed during the accretion/merger process, in contrast
to the gaseous merger models, and the total number of globular clusters and their metallicities
are determined by the mass of the proto-galactic fragment where globular clusters formed.

Beasley et al. (2002) presented a simulation of the globular cluster systems of 450 elliptical galaxies
in the $\Lambda$CDM model, assuming that the globular clusters form at two epochs,
first at $z>5$ in protogalactic fragments, and second during the subsequent gas-rich
merging of these fragments. They found
1) the GCs in most galaxies  show bimodality in metallicity,
2) the RGCs (9 Gyr) are on average younger than the BGCs (12 Gyr),
3) the RGCs show a large range in age (5 to 12 Gyr), which increases for low-luminosity galaxies,
and for galaxies with low circular velocity halos, and
4) the RGCs formed via gaseous merging, the bulk of which occurs at $1<z<4$.

A few remaining problems in this model are
1) their model galaxies do not show the well-known color-magnitude relation of elliptical galaxies
(see their Figure 12),
2) they had to truncate the formation of the BGCs at $z\approx 5$ to avoid   too high
metallicity at $z=0$, the origin of which needs to be explained,
and
3) they used formation efficiency of the RGCs, 0.7\%. % and and the BGCs 0.2\%.
This value for the RGCs is 3-4 times higher than those for M87 and M49 (C\^ote 2002).

Kravtsov \& Gnedin (2003) found, in their numerical simulation for the formation of globular clusters
in the Milky Way-size galaxy, that the best conditions for globular
cluster formation at the high density core of the supergiant molecular clouds  in the
gaseous disks of the proto-galaxy
are at $z\sim 3-5$, although the first globular clusters start to appear at $z\approx 12$.
Their simulation covered $z>3$.

C\^ote (2002) pointed out that various names (proto-galactic fragment, supergiant molecular
clouds, proto-galactic disks, Searle-Zinn fragments, and failed dwarfs) are the same
objects in their physical properties.

\subsection{Summary}

There are several models which can explain well several observational constraints, but none of
them is free from a few major limits at the moment. However, it appears that the concept
`simple is beautiful' is changing to `complexity and delicacy are reality'.
Finally we may need a bibimbap model to explain the formation of giant elliptical
galaxies and globular clusters (bibimbap is a very delicious Korean rice mixed with various
vegetables and some meat, and bibim (Korean) means mixing several).

\section{CONCLUSION AND FUTURE}

\begin{figure*}
 \begin{minipage}{\textwidth} \begin{center}
    \leavevmode
    \epsfxsize = 16.0cm
    \epsfysize = 16.0cm
    \epsffile{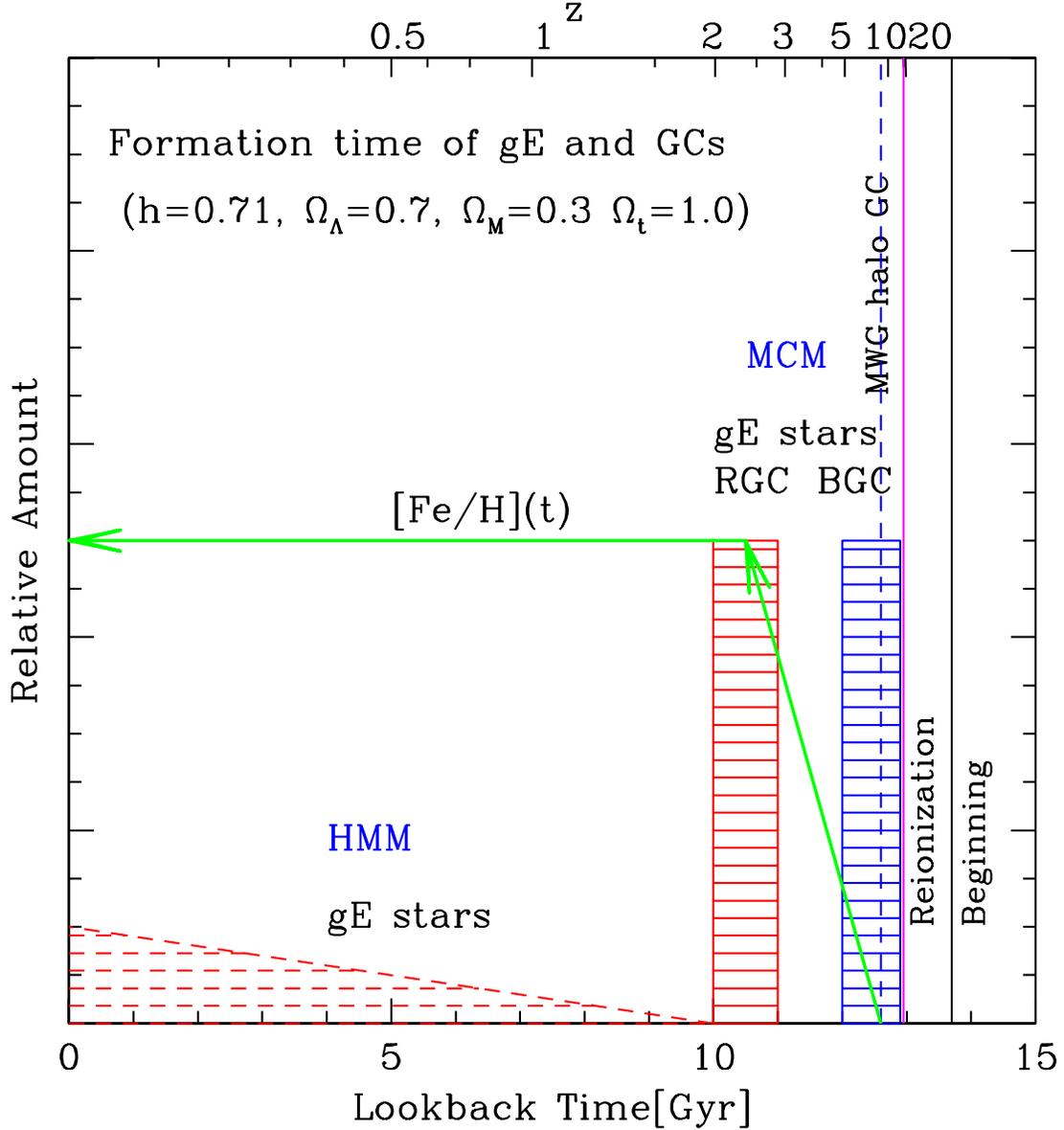} %Eformtime.ps}
  \end{center}
  \vspace{-0.2in}
\caption{A schematic diagram showing the formation time of gEs, BGCs and RGCs 
according to
the Monolithic Collapse Models (MCM) and the Hierarchical Merging Models (HMM).
Lookback time($z$) is based on the concordance cosmological model.
Magenta solid line at the lookback time 12.95 Gyr ($z=17$) represents the 
reionization epoch,
indicating the birth of the first stellar systems.
Blue dashed line at the lookback time 12.6 Gyr represents the mean age of the 17
 metal-poor
globular clusters in the halo of our Galaxy.
Green arrow line represents a sketch for chemical enrichment history needed.
}
\label{fig14} \end{minipage}
\end{figure*}

Most giant elliptical galaxies show a bimodal color distribution of globular clusters,
indicating a factor of $\approx 20$ metallicity difference between the two.
The RGCs (metal-rich globular clusters)
are closely related with the stellar halo in color and spatial distribution,
while the BGCs (metal-poor globular clusters) are not.
The ratio of the number of the RGC and the BGC varies depending on
galaxies.
Considering the observational constraints from stars and globular clusters in giant elliptical
galaxies,
I draw the following conclusive constraints
for understanding when, how long, and how elliptical
galaxies and globular clusters formed, which are sketched in Figure 14.

\begin{enumerate}

\item{There are found to be many giant elliptical galaxies which formed at $z>2$ ($>10$ Gyr ago).}

\item{The RGCs and most stars in elliptical galaxies formed at the similar time $>10$ Gyr ago.}

\item{The BGCs formed before the RGCs, but after reionization at $z \approx 10-30$ ($<13$ Gyr ago).}

\item{The BGCs all must have formed out of metal-poor clouds, independently of most stars 
in their present host galaxy.}
%They formed elsewhere around the similar time, and were captured later by the present host galaxies.}

\item{There were no major bursts between the RGCs and the BGCs,
but the globular clusters continued forming around the epochs of the BGC and RGC peaks.}

\item{Therefore there must have been two major bursts
with a gap during the period of 10--13 Gyr ago,
 and the formation duration  should  be $\approx 1$ Gyr for the BGCs,
and  $\approx 1$ Gyr or little longer for the RGCs.
The major epochs for the formation of the BGCs and the RGCs may be
12.5 Gyr and 10.5 Gyr (presented by the blue and red bars in Figure 14).
Note that 12.5 Gyr is similar to the mean age of the metal-poor globular
clusters in the Galactic halo.}

\item{From the epoch of the BGC formation to that of the RGC formation, the metallicity
should be increased by a factor of 20 ($\Delta {\rm [Fe/H]} = 1.0$ dex).}

\item{Both the MCM and HMM are needed, but the HMM needs to have a way to make giant elliptical
galaxies form at $>10$ Gyr ago (much earlier than the present HMMs predict).
The MCM by itself cannot reproduce the large-scale structures.}

\item{Merging (major, minor) should continue to occur, while keeping the color-magnitude
relations, the fundamental planes, and bimodality of globular clusters.}

%\item{These exist halo stars with metallicity lower than the halo globular clusters in our Galaxy.
%Then some metal-poor stars formed before or at the same time as the BGC. There exist a small
%number of metal-poor stars in NGC 5128, showing some stars formed before the RGCs.}

\end{enumerate}

Some remaining problems are:

\begin{enumerate}

\item{Why there is a gap between the BGC formation epoch and the RGC formation epoch, seen over about 5
magnitude range of galaxies (by a factor of 100 in the total luminosity)?}

\item{How did the metallicity increase by a factor of 20
during the short gap ($\Delta t \approx 2$ Gyr)?}

\item{Why are kinematics of the globular clusters in M49, M87 and NGC 1399  so different?
Is it just due to a statistical effect or due to intrinsic difference?}

\item{How can the HMM make giant elliptical galaxies  $>10$ Gyr ago?}

\end{enumerate}

A list for the future is:

\begin{enumerate}
\item{Determination of the relative age of the BGCs and RGCs with an accuracy of $\approx 1$ Gyr
is desperately needed to solve the unknown mystery directly.
For this we need to reduce the measurement errors of spectral lines
or photometric colors (optical and near-IR), and to design new ways of precise
age estimates. At the same time, we need to improve isochrone models based on population synthesis.}

\item{We need to understand the diversity in the kinematics of the globular clusters
in M87, M49 and NGC 1399, and need to investigate the kinematics of the globular clusters
in more giant elliptical galaxies.}

\item{Search for intracluster globular clusters and intragalactic globular cluster may produce
an interesting result .
There are some reports related with this issue (Hilker 2002, Jord\'an et al. 2002b), but more are needed.}

\item{Ultimatum models to explain both stars and globular clusters in galaxies are needed.
When will it be?}
\end{enumerate}

\vskip 0.4cm

The author is grateful to his collaborators on the project of extragalactic globular clusters,
especially, Doug Geisler, Ata Sarajedini, Brad Whitmore, Rupali Chandar, Nobuo Arimoto,
Eunhyeuk Kim, Sang Chul Kim, Hong Soo Park and Ho Seong Hwang,
and to Prof.Hyung Mok Lee for inviting him to the exciting meeting on the
formation and interaction of galaxies during the Korean heatwave of Worldcup soccer 2002.
The author thanks the staff of
the Department of Terrestrial Magnetism, Carnegie Institution of Washington,
for their kind hospitality while writing this paper as Visiting Investigator.
Vera Rubin is thanked for her kind and careful reading the draft, for pointing out an important
reference by Baum (1959), and for showing what an astronomer is.
This project was supported in part by the Korean Research Foundation Grant
(KRF-2000-DP0450) and the BK21 program.

%\section{REFERENCES}

\end{document}